\documentclass[12pt, dvipdfmx]{article}
\usepackage{enumerate}
\usepackage{caption}
\usepackage{amsmath}	
\usepackage{amssymb}
\usepackage{epic}
\usepackage{epic,eepic}
\usepackage{fancyheadings}
\usepackage[dvips]{graphicx}
\usepackage{graphics}
\usepackage{graphicx}
\usepackage{latexsym}
\usepackage{amssymb}
\usepackage{ulem}
\usepackage{xspace}
\usepackage{lscape}
\usepackage{yhmath}
\usepackage[top=30truemm, bottom=30truemm, left=25truemm, right=25truemm]{geometry}
\usepackage{bm}
\usepackage{xcolor}
\allowdisplaybreaks

\usepackage{delarray}

\def\IntKer2{R(K)}

\newtheorem{Theorem}{Theorem}

\newtheorem{Remark}{Remark}

\usepackage[authoryear]{natbib}
\usepackage[dvipdfmx, colorlinks=true, linkcolor=black, citecolor=black, urlcolor=black]{hyperref}
\begin{document}


\vspace{5mm}
\title{Sparse Variable Sharpening in High-Dimensional Kernel Density Estimation}
\vspace{5mm}
\date{}
\maketitle
\begin{center}
Kiheiji NISHIDA
\footnote{Department of Business Administration, Kyoto Sangyo University. Address: 2K539, Kamigamo-Motoyama, Kita-ku, Kyoto, 603-8555, JAPAN. E-mail: kiheiji.nishida@gmail.com}
\end{center}
\vskip 10mm
\noindent {\bf{ABSTRACT}}\\
High-dimensional kernel density estimation suffers from the curse of dimensionality. This study proposes a hybrid density estimator defined as the product of a joint density over a pre-specified subset of dimensions and marginal densities for the remaining variables, instead of estimating the full high-dimensional density directly. Under this framework, given the observed data, we select a subset of variables for joint density construction to improve estimation accuracy, while modeling the remaining variables through their respective marginal densities. This construction involves a trade-off between the approximation error induced by simplifying the dependence structure and the variance reduction achieved by lowering the dimension of the joint density. We employ a genetic algorithm to efficiently identify such variable subsets. Theoretical and numerical results demonstrate that the proposed estimator can outperform conventional full-dimensional kernel density estimation when this trade-off is appropriately balanced.
\section{Introduction}

Suppose that $\{ \mathbf{X}_{i} \}_{i=1}^{N}$ is a $d$-dimensional i.i.d. sample of size $N$ generated from the true density function $f(\mathbf{x})$ on $\mathbb{R}^{d}$, where $\mathbf{x}^{\top} = (x_{1}, x_{2}, ..., x_{d})$ and $\mathbf{X}_{i}^{\top}=(X_{i1}, X_{i2}, ..., X_{id})$, and let $\mathbf{X}$ denote the corresponding $N \times d$ data matrix. Our aim is to estimate $f(\mathbf{x})$ using the multivariate kernel density estimator (KDE) written as
\begin{equation*} 
\widehat{f}_{\mathbf{H}}(\mathbf{x}) = \frac{1}{N} \sum_{i=1}^{N} K_{\mathbf{H}}(\mathbf{x} - \mathbf{X}_{i}),
\end{equation*}
where $\mathbf{H}$ is a symmetric and positive definite $d$-dimensional bandwidth matrix and $K_{\mathbf{H}}(\mathbf{t}) = |\mathbf{H}|^{-\frac{1}{2}}K(\mathbf{H}^{-\frac{1}{2}} \mathbf{t})$ is a non-negative real valued bounded kernel function scaled by $\mathbf{H}$. Under the assumptions that $\int K(\mathbf{t}) d \mathbf{t} = 1$, $\int \mathbf{t} K(\mathbf{t}) d \mathbf{t} = 0$, and $\int \mathbf{t} \mathbf{t}^{\top} K(\mathbf{t}) d \mathbf{t} = \mu_{2}(K) \mathbf{I}_{d}, \mu_{2}(K) > 0$, the pointwise convergence rate is given by $O\bigl( N^{-\frac{4}{d+4}} \bigr)$. It is well known that as the dimension increases, the convergence rate deteriorates, leading to the so-called curse of dimensionality. In particular, the impact of dimensionality arises primarily from the variance term, whose order deteriorates exponentially with the dimension, whereas the bias term does not explicitly depend on the dimension for a fixed bandwidth.

Several approaches have been proposed to overcome the curse of dimensionality. One such method is the reduced set density estimator (Girolami and He \citeyear{Girolami.2003}; He and Girolami \citeyear{He.2004}), which selects a subset of the original sample to minimise the estimation error of the resulting KDE. Since the curse of dimensionality arises from the fact that the effective sample size per evaluation point decreases with increasing dimensionality, selecting an appropriate small subset can be beneficial for estimating high-dimensional KDEs. Nishida (\citeyear{Nishida.2023}, \citeyear{Nishida.2026}) also proposes a data condensation method for the same purpose via a genetic algorithm (GA) (e.g., Goldberg \citeyear{Goldberg.1988}; Goldberg \citeyear{Goldberg.1989}; Holland \citeyear{Holland.1992}; Forrest \citeyear{Forrest.1993}; Haupt \citeyear{Haupt.2004}). Related approaches include optimal transport-based subsampling methods (Zhang et al., \citeyear{Zhang.2023}), which select a representative subset to approximate the original KDE. Another approach avoids the curse of dimensionality by decomposing a multivariate function into a sum of univariate components. Additive models (e.g., Hastie and Tibshirani, \citeyear{Hastie.1990}) approximate a high-dimensional function through a collection of one-dimensional functions, thereby reducing the estimation problem to a set of univariate tasks. However, such models cannot capture interactions among variables when strong dependencies are present. 
Related approaches include projection-based methods such as projection pursuit density estimation (PPDE) (Friedman et al., \citeyear{Friedman.1984}; Huber, \citeyear{Huber.1985}), which reduce the effective dimensionality by projecting the data onto lower-dimensional subspaces and model a high-dimensional density through a sequence of one-dimensional projections, thereby alleviating the curse of dimensionality. However, PPDE relies on the estimation of projection directions and may suffer from optimization instability and limited interpretability, since the projection directions are linear combinations of the original variables, making it difficult to interpret the contribution of individual variables. 
Related approaches include decomposable graphical models (Schwaighofer et al., \citeyear{Schwaighofer.2007}) and forest-structured density estimation (Liu et al., \citeyear{Liu.2011}).

In this study, rather than performing a straightforward $d$-dimensional estimation, we avoid confronting the curse of dimensionality directly. Let $S \subseteq \{1,2,...,d \}$ denote a subset of variable indices called active index set, and let $\mathbf{x}_{S}^{\top} =(x_j)_{j \in {S}}$ and $\mathbf{x}_{\bar{S}}^{\top} =(x_j)_{j \notin {S}}$ be the corresponding subvectors of $\mathbf{x}^{\top} = (x_1,\ldots,x_d)$. For a given $S$, let $f_{S}(\mathbf{x}_{S})$ denote a $|S|$-dimensional joint density function consisting of all the variables in $S$ and let $f_{j}(x_{j})$ denote the marginal density of $f(\mathbf{x})$ with respect to $x_{j}$. We consider the restricted factorized approximation : 
\begin{eqnarray} \label{def_F}
{q}_{S}(\mathbf{x}) = {f_{S}}(\mathbf{x}_{S})\prod_{j \notin S} {f_{j}}(x_{j}), \ \ \ |S| \le b,
\end{eqnarray}
where the tuning parameter $b \le d$ denotes the maximum number of variables allowed in the joint distribution. Then, we define the population-optimal active index set within this family as
\begin{eqnarray} \label{opt.S}
S_{0} = \arg \min_{S : |S| \le b} \int_{\mathbb{R}^{d}} [{q}_{S}(\mathbf{x}) - f(\mathbf{x})]^{2} d\mathbf{x} \nonumber
\end{eqnarray}
and denote the corresponding optimal approximation by $q_{0}(\mathbf{x}) = q_{S_{0}}(\mathbf{x})$. The problem thus reduces to how to efficiently select the active index set $S$ to model the joint density from the $d$ available variables--while treating the remaining variables through their marginal densities--so that the resulting approximation closely approximates the true density $f$.

If we employ kernel approach and let $\boldsymbol{\tau} \in \mathbb{Z}_{\ge 0}^{d}$ denote an integer-valued $d$-dimensional vector, the resulting estimator, hereafter called the sparse variable-sharpening KDE (SVS-KDE), or simply SVS, is given by
\begin{eqnarray} \label{VS.KDE}
{\widehat{q_{S}}(\mathbf{x})}
&=& \Biggl[ \frac{1}{N} \sum_{i=1}^{N} \prod_{j \in S} \frac{K_{h_{S}} \Bigl( x_{j} - X_{ij} \Bigr)^{\tau_{j}}}{C_{h_{S}}(\tau_{j})} \Biggr] \Biggl[ \prod_{j \notin S} \Biggl\{ \frac{1}{N} \sum_{i=1}^{N} K_{h_{j}} \Bigl( x_{j} - X_{ij} \Bigr) \Biggr\} \Biggr], \\
& & \nonumber \\
\mbox{subject to\ } & & \tau_{j} \in \{0, 1, 2, 3, \ldots, b \}, \ \ \ \sum_{j \in S} \tau_{j} = b, \nonumber \\
& & S = S(\boldsymbol{\tau}) \equiv \bigl \{j \bigl| \tau_{j} > 0, \ j \in \{1,2,...,d \} \bigr \}, \nonumber
\end{eqnarray}
where $(S, h_{S})$ should be optimized, given the observed data, so as to approximate the true density $f$ as accurately as possible. Here, we employ the scalar bandwidth matrix $\mathbf{H}_{S} = h_{S}^{2} \mathbf{I}_{|S|}$, and $C_{h}(\tau_{j}) = \int K_{h}(t)^{\tau_{j}} dt$ denotes the normalizing constant for $\tau_{j} \ge 1$. Furthermore, $h_{k}$ is the bandwidth of the marginal KDE for the variable $x_{k}$. The parameter $\tau_{j}$ represents the degree of sharpness of the distribution along the corresponding coordinate direction. This interpretation becomes clear when a Gaussian product kernel is employed in \eqref{VS.KDE}, in which case the kernel of the joint part is given by:
\begin{eqnarray} \label{Gaussian.product.ker}
\prod_{j \in S} \frac{K_{h_{S}}(x_{j} - X_{ij})^{\tau_{j}}}{C_{h_{S}}(\tau_{j})} = \frac{1}{\prod_{j \in S} \sqrt{2\pi \bigl(\frac{h_{S}^2}{\tau_{j}} \bigr)}} \exp \Biggl[ - \sum_{j \in S} \frac{\bigl(x_{j} - X_{ij} \bigl)^2}{2 \bigl( \frac{h_{S}^2}{\tau_{j}} \bigr)} \Biggr],
\end{eqnarray}
where term $h_{S}^2/\tau_{j}$ can be interpreted as the variance of the kernel along the $j$-th axis. Under the constraint of $\sum_{j \in S} \tau_{j} = b$, the problem can also be interpreted as determining how the total sharpness $b$ should be allocated among the variables used in the joint density estimation.

The advantage of this approach is that, instead of estimating a full $d$-dimensional KDE, we only need to estimate $|S|$-dimensional KDE together with $d-|S|$ one-dimensional KDEs. As a result, when the sample size is insufficient for reliable full-dimensional estimation, this method may yield a more accurate approximation than the standard $d$-dimensional KDE. 
This phenomenon can be attributed to the following two reasons. First, from the perspective of degrees of freedom, the $d$-dimensional joint density estimated using the conventional approach involves only a single bandwidth parameter, whereas the proposed method has $d - |S| + 1$ degrees of freedom. Second, the asymptotic mean integrated squared error behavior differs across the competing estimators in infinite sample setting. For the proposed estimator $\widehat{q_{S}}(\mathbf{x})$, which combines a $|S|$-dimensional joint component with $d-|S|$ marginal components, the asymptotic mean integrated squared error (AMISE) satisfies, for any fixed $0 < \varepsilon < 1$,
\begin{eqnarray} \label{MISE.SVS}
\lefteqn{(1 - \varepsilon)\mathcal{E}(f,q_{S}) - O \left(N^{-\frac{4}{|S|+4}}+|\bar S|N^{-\frac45} \right)} \nonumber \\
&\le& \mbox{AMISE}\left(f(\mathbf{x}),\widehat{q}_S(\mathbf{x})\right) \le (1 + \varepsilon) \mathcal{E}(f,q_{S}) + O \left(N^{-\frac{4}{|S|+4}} + |\bar S| N^{-\frac{4}{5}} \right),
\end{eqnarray}
where $\mathcal{E}(f, q_{S})$ represents the approximation error $\mathcal{E}(f, q_{S}) = \int_{\mathbb{R}^{d}} [f(\mathbf{x}) - q_{S}(\mathbf{x})]^2 d\mathbf{x}$, which is independent of the sample size $N$.
As $N \to \infty$, the estimation error vanishes, and since $\varepsilon$ can be chosen arbitrarily small, the AMISE converges to the approximation error. If $q_{S}(\mathbf{x})$ is identical to the true density $f(\mathbf{x})$, the approximation error $\mathcal{E}(f, q_{S})$ vanishes, and hence the AMISE converges to zero. On the other hand, when $q_S(\mathbf{x})$ differs from $f(\mathbf{x})$, the SVS estimator has a nonzero asymptotic approximation error, whereas the full joint KDE has no such approximation error but has the optimal AMISE rate $O(N^{-\frac{4}{d+4}})$. Therefore, depending on the configuration of $d$, $S$, $N$, and $f$, it is possible that the proposed estimator may outperform the full joint KDE for finite $N$.

This advantage can be understood as a trade-off between approximation error and variance reduction with respect to $|S|$. As $|S|$ increases, the dimension of the joint part becomes larger, which improves the approximation capability of $q_{S}(\mathbf{x})$ and hence reduces the approximation error $\mathcal{E}(f, q_{S})$. However, at the same time, the variance increases due to the curse of dimensionality. After bandwidth optimization, this effect is reflected in the term $O \Bigl( N^{-\frac{4}{|S| + 4}} \Bigr)$ in (\ref{MISE.SVS}). Conversely, when $|S|$ is small, the variance is reduced because the effective dimensionality of the joint estimation is lower, but this comes at the cost of a larger approximation error $\mathcal{E}(f, q_{S})$. Therefore, the performance of the proposed estimator critically depends on the choice of $S$, which balances the variance reduction and the approximation accuracy. This suggests that the optimal choice of $S$ depends on the underlying structure of $f(\mathbf{x})$ and the sample size $N$, and should be determined in a data-adaptive manner. However, since deriving an analytical method to identify such a subset $S$ is generally difficult, we propose in Section~3 a metaheuristic algorithm to identify $S$, extending the GA introduced in Nishida (2023, 2026). The parameter $b$ in \eqref{def_F} is introduced to the GA as a device to facilitate the search for $S$, by controlling the maximum size of the joint component, to be explained in Section~\ref{alg}.

The remainder of this paper is organized as follows. In Section~\ref{theoretical.results}, we derive the mean, variance, AMISE, and asymptotic normality of SVS for a fixed active index set and sharpening configuration. The proofs of the theorems are given in the \hyperref[Appendix]{Appendix}. In Section~\ref{alg}, we present the algorithm to estimate SVS using GA. To validate our proposed estimator, we give the numerical experiments in Section~\ref{numerical.experiments} and the real data application in Section~\ref{real.data}. Section~\ref{discussion} discusses the results of the study.
\section{Theoretical results} \label{theoretical.results}

Throughout this section, the active index set $S$ and the sharpening parameters $\tau$ are treated as fixed and non-random. The bandwidths are allowed to depend on $N$ through the asymptotic conditions stated below. Accordingly, the following results characterize the SVS estimator for a given $(S,\tau)$, rather than the fully data-driven estimator obtained after LSCV and GA selection. The mean and variance of $\widehat{q_{S}}(\mathbf{x})$ are given in Theorems~\ref{Theorem.E.SVS} and~\ref{Theorem.Var.SVS}, respectively. Bounds for the AMISE and the bandwidth-optimized estimation-error rate are given in Theorem~\ref{Theorem.MISE.bound}, and asymptotic normality is established in Theorem~\ref{Theorem.AN.SVS}
. We denote the term $R(f)$ to be the integral of the squared function $f(\cdot)$. We denote the kernel for the joint part to be
\begin{eqnarray}
K_{S}(\mathbf{t}_{S}) = \prod_{j \in S} \frac{K(t_{j})^{\tau_{j}}}{C(\tau_{j})}. \nonumber
\end{eqnarray}
For notational convenience, we also define
\begin{eqnarray}
W_{B} = h_{S}^{4} + \sum_{j \in \bar S} h_{j}^{4}, \ \ \ W_{V} = \frac{1}{N h_{S}^{|S|}} + \sum_{j \in \bar S} \frac{1}{N h_{j}}, \ \ \ and\ \ W_{N} = W_{B} + W_{V}. \nonumber
\end{eqnarray}
Then, we place the following assumptions throughout the proofs. \\\\
{\bf{A1.}} The bandwidth matrix for the joint part is $\mathbf{H}_{S} = h_{S}^{2} \mathbf{I}_{S}$ and satisfies $h_{S} \to 0$, $N h_{S}^{|S|} \to \infty$ as $N \to \infty$.\\
{\bf{A2.}} The bandwidths for the marginal components satisfy $h_{j} \to 0$, $N h_{j} \to \infty, j \in \bar{S}$ as $N \to \infty$. \\
{\bf{A3.}} The density $f(\mathbf{x})$ and all its marginal densities are bounded. The densities $f_{S}(\mathbf{x}_{S})$ and $f_{j}(x_{j})$, $j \in \bar{S}$, are strictly positive and twice continuously differentiable, and their second-order derivatives are square-integrable.\\
{\bf{A4.}} \noindent
The kernels $K_S$ and $k$ are bounded and absolutely integrable with unit integral, zero first moments, and finite second moments and squared integrals. For each $j \in S$, define the coordinate-specific second moment by
\begin{eqnarray}
\mu_{2,j}(K_{S}) = \int_{\mathbb{R}^{|S|}} t_{j}^{2} K_{S}(\mathbf{t}_S) d\mathbf{t}_{S}. \nonumber
\end{eqnarray}
{\bf{A5.}} For $|S| > 1$, the bandwidths $h_{S}$ and $h_{j}$, $j \in \bar{S}$, satisfy
\begin{eqnarray}
h_{S} \to 0, \ \ \ N h_{S}^{|S|} \to \infty, \nonumber
\end{eqnarray}
and
\begin{eqnarray}
h_{j} \to 0, \ \ \ \frac{h_{S}^{|S|}}{h_{j}} \to 0, \ \ \ as \ N \to \infty.  \nonumber
\end{eqnarray}
\medskip
Theorems~\ref{Theorem.E.SVS}-\ref{Theorem.AN.SVS} and Remarks~\ref{Remark.consist}-\ref{Remark.N.cross.over}
 are given below. All the proofs of the theorems are given in the \hyperref[Appendix]{Appendix}.
\medskip
{{\Theorem{\hspace{-2.0mm}{\bf{.}} \label{Theorem.E.SVS} \hspace{-0.1mm} Under assumptions {\bf{A1-4}}, the mean of $\widehat{q_{S}}(\mathbf{x})$ is given as follows:
\begin{eqnarray} \label{E.SVS}
\lefteqn{E \Bigl[ \widehat{q_{S}}(\mathbf{x}) \Bigr] = q_{S}(\mathbf{x}) + \frac{h_{S}^{2}}{2} \Biggl[ \sum_{j \in S} \mu_{2, j}(K_{S}) \frac{\partial^{2} f_{S}(\mathbf{x}_{S})}{\partial x_{j}^{2}} \Biggr] \prod_{k \in \bar{S}} f_{k}(x_{k})} \nonumber \\
& & + \frac{\mu_{2}(k)}{2} f_{S}(\mathbf{x}_{S}) \Biggl[ \sum_{k \in \bar{S}} h_{k}^{2} f_{k}^{(2)}(x_{k}) \prod_{l \in \bar{S}, l \neq k} f_{l}(x_{l}) \Biggr] + o \Biggl( h_{S}^{2} + \sum_{k \in \bar{S}} h_{k}^{2} \Biggr) + O\Bigl(\frac{1}{N} \Bigr). \nonumber
\end{eqnarray}
}}
\medskip
{{\Theorem{\hspace{-2.0mm}{\bf{.}} \label{Theorem.Var.SVS} \hspace{-0.1mm} Under assumptions {\bf{A1-4}}, the variance of $\widehat{q_{S}}(\mathbf{x})$ is given as follows:
\begin{eqnarray} \label{V.SVS}
\mathrm{Var}\Bigl[ \widehat{q_{S}}(\mathbf{x}) \Bigr] = q_{S}(\mathbf{x})^2 \Biggl[ \frac{R(K_{S})}{Nh_{S}^{|S|}f_{S}(\mathbf{x}_{S})} + \sum_{j \in \bar{S}} \frac{R(k)}{Nh_{j}f_{j}(\mathbf{x}_{j})} \Biggr] + o \Biggl( \frac{1}{Nh_{S}^{|S|}} + \sum_{j \in \bar{S}} \frac{1}{Nh_{j}} \Biggr). \nonumber
\end{eqnarray}
}}
\medskip
{{\Theorem{\hspace{-2.0mm}{\bf{.}} \label{Theorem.MISE.bound} \hspace{-0.1mm} Let $LB(\mathbf{x})$ and $LV(\mathbf{x})$ denote the leading bias and variance terms of $\widehat{q_{S}}(\mathbf{x})$ in Theorems~\ref{Theorem.E.SVS} and \ref{Theorem.Var.SVS}, respectively. Define the approximation error and the AMISE, respectively, by
\begin{eqnarray}
\mathcal{E}(f,q_{S}) = \int_{\mathbb{R}^d} \bigl( q_{S}(\mathbf{x}) - f(\mathbf{x}) \bigr)^2 d\mathbf{x} \nonumber
\end{eqnarray}
and
\begin{eqnarray}
{AMISE} \bigl(\widehat{q_{S}}(\mathbf{x}), f(\mathbf{x}) \bigr) = \int_{\mathbb{R}^d} \bigl( q_{S}(\mathbf{x}) - f(\mathbf{x} \bigr) + LB(\mathbf{x}) \bigr)^2 d\mathbf{x} + \int_{\mathbb{R}^d} LV(\mathbf{x})d\mathbf{x}. \nonumber
\end{eqnarray}
Under assumptions {\bf{A1-4}}, for any fixed $0 < \varepsilon < 1$, 
\begin{eqnarray}
(1 - \varepsilon) \mathcal{E}(f, q_{S}) - O(W_{B}) \leq {AMISE} \bigl( \widehat{q_{S}}(\mathbf{x}), f(\mathbf{x}) \bigr) \leq (1 + \varepsilon) \mathcal{E}(f,q_{S}) + O(W_{N}). \ \ \ \label{MISE_bound}  \nonumber
\end{eqnarray}
After bandwidth optimization, $W_{N}$ and $W_{B}$ satisfy
\begin{eqnarray}
W_{B}, W_{N} = O \left( N^{-\frac{4}{|S|+4}} + |\bar S| N^{-\frac{4}{5}} \right). \nonumber
\end{eqnarray}}}
\medskip
{\Remark{\hspace{-2.0mm}{\bf{.}}} \label{Remark.consist}} Theorem~\ref{Theorem.MISE.bound} shows that the AMISE consists asymptotically of the approximation error. In particular, as $N \to \infty$, both $W_{B}$ and $W_{V}$ vanish, and hence, for any fixed $0 < \varepsilon < 1$,
\begin{eqnarray}
(1 - \varepsilon) \mathcal E(f,q_{S}) \le \liminf_{N \to \infty} \operatorname{AMISE} \le \limsup_{N \to \infty} \mbox{AMISE} \le (1 + \varepsilon) \mathcal E(f,q_{S}). \nonumber
\end{eqnarray}
Since $\varepsilon$ can be chosen arbitrarily small, it follows that
\begin{eqnarray}
\mbox{AMISE} \to \mathcal{E}(f, q_{S}), \ \ \text{as}\ N \to \infty. \nonumber
\end{eqnarray}
Thus, for a fixed active index set $S$, the approximation error $\mathcal{E}(f,q_{S})$ represents the asymptotic error floor of the AMISE. In particular, if $q_{S} = f$, then $\mathcal{E}(f,q_{S}) = 0$, and hence the AMISE converges to zero. 
\medskip
{\Remark{\hspace{-2.0mm}{\bf{.}}} \label{Remark.N.cross.over}} For a finite sample size, the choice of $S$ involves a trade-off between approximation and estimation errors. Increasing $|S|$ reduces $\mathcal{E}(f,q_{S})$ by retaining more of the dependence structure of $f$, but increases the estimation error of the joint KDE. Conversely, reducing $|S|$ lowers the estimation error, primarily through variance reduction resulting from the lower-dimensional joint KDE, at the cost of a larger approximation error. Thus, the choice of $S$ may depend on the sample size $N$. This trade-off also explains when SVS can outperform the full-dimensional KDE. For sufficiently large $N$, Theorem~\ref{Theorem.MISE.bound} suggests $\mbox{AMISE}_{\mathrm{SVS}} \to \mathcal{E}(f,q_{S})$, whereas the optimally tuned full $d$-dimensional KDE has $\mbox{AMISE}_{\mathrm{full}} \approx C_{\mathrm{full}} N^{-\frac{4}{d+4}}$. The estimation error of SVS decreases relatively rapidly due to its lower-dimensional structure, whereas the full-dimensional KDE converges more slowly but has no approximation error. Hence, an approximate crossover sample size is
\begin{eqnarray}
N^{*} \approx \left[ \frac{C_{\mathrm{full}}}{\mathcal{E}(f,q_{S})} \right]^{\frac{d+4}{4}}. \nonumber
\end{eqnarray}
Thus, a smaller approximation error leads to a larger $N^{*}$. Moreover, provided that $C_{\mathrm{full}}/\mathcal{E}(f,q_{S})>1$, the approximation suggests that $N^{*}$ increases with $d$. If $E(f,q_{S}) = 0$, SVS has no asymptotic approximation error, and the crossover formula for $N^{*}$ does not apply.
\medskip
{{\Theorem{\hspace{-2.0mm}{\bf{.}} \label{Theorem.AN.SVS} \hspace{-0.1mm} Suppose that assumptions {\bf{A1-5}} are satisfied. For a fixed active set $S$ with $|S| > 1$ and each fixed $\mathbf{x}$,
\begin{eqnarray}
\sqrt{N h_{S}^{|S|}} \biggl( \widehat{q_{S}}(\mathbf{x}) - q_{S}(\mathbf{x}) - \mathrm{Bias} \left(\widehat{q_{S}}(\mathbf{x}) \right) \biggr)
\overset{d}{\to} N \Biggl(0,\ \biggl( \prod_{k \in \bar{S}} f_{k}(x_{k}) \biggr)^{2} f_{S}(\mathbf{x}_{S}) R(K_{S}) \Biggr). \nonumber
\end{eqnarray}}}
\section{Algorithm} \label{alg}

In practice, $S$, $\tau$, and the joint bandwidth are unknown and are selected from the observed data using the GA described below. In the context of GA, the term {\it{fitness}} is used to indicate that larger values are more desirable; therefore, a fitness function is defined as the negative of the error. Let $V(\boldsymbol{\tau}, \mathbf{H}_{S(\boldsymbol{\tau})})$ be the fitness function evaluating the SVS in \eqref{VS.KDE}, written as $\widehat{q_{\mathbf{H}_S(\boldsymbol{\tau})}}(\mathbf{x}) \equiv \widehat{q_{S, \mathbf{H}_S(\boldsymbol{\tau})}}(\mathbf{x})$, where the bandwidth matrix for the joint part is $\mathbf{H}_S = h_S^{2}\mathbf{I}_d$. The bandwidths for the marginal components are computed prior to the algorithm and remain fixed throughout. Define $\mathbf{D} = \{\iota_1, \iota_2, \ldots, \iota_b\}$ as the set of variable indices, where $\iota_i \in \{1,2,\ldots,d\}$ for $i=1,\ldots,b$ and $|\mathbf{D}| = b$; repetitions are allowed, so $\mathbf{D}$ is a multiset. For example, when $d=10, b=5$, $\mathbf{D}=\{1,1,2,2,4 \}$, we obtain $\boldsymbol{\tau} = \{2,2,0,1,0,0,0,0,0,0 \}$ and $S(\boldsymbol{\tau})=\{1,2,4\}$. A GA is employed to search for an optimal solution $(S(\boldsymbol{\tau}^{*}), \mathbf{H}^{*})$, as the fitness function is not necessarily convex. In the subsequent algorithm, for notational simplicity, we write $S(\boldsymbol{\tau}_{i})$ instead of $S(\boldsymbol{\tau}_{i}^{(g-1)})$ whenever the generation index is clear from the context. \\\\
{\bf{The proposed GA}}
\\\\{\bf{Step 0:}} Compute the kernel marginal density function for each variable $j = 1,2,...,d$, denoted by $\widehat{f_{j}}(x_{j})$.
\\\\{\bf{Step 1:}} Initial generation $g=1$:
\begin{enumerate}
\item Set the size of the index subset to $b$, the number of index subsets to $B$, and the total number of iterations to $G$.
\item Make $B$ index subsets, each of size $b$, by sampling with replacement from the original index set $\{1,2,\ldots,d\}$, where $B$ is an even number. Each subset is called {\it{chromosome}} and is denoted as $\mathbf{D}_{i}^{(1)} = \{\iota_{i, 1}^{(1)}, \iota_{i, 2}^{(1)}, ..., \iota_{i, b}^{(1)} \}$, $i=1,2,...,B$, where $\iota_{i, j}^{(1)}$, $j=1,2,...,b$, is the $j$-th index of the $i$-th index subset called {\it{gene}}. The {\it{population}} in generation~1 is written as $\mathbf{D}^{(1)} = \{\mathbf{D}_{1}^{(1)}, \mathbf{D}_{2}^{(1)}, ..., \mathbf{D}_{B}^{(1)} \}$.
\end{enumerate}
{\bf{Step 2:}}
The generations $g=2,3,...,G-1$:
\begin{enumerate}
\item Inherit population $\mathbf{D}^{(g-1)} = \{\mathbf{D}_{1}^{(g-1)}, \mathbf{D}_{2}^{(g-1)}, ..., \mathbf{D}_{B}^{(g-1)} \}$ from the previous generation $g-1$.
\item For each subset $\mathbf{D}_{i}^{(g-1)}$, $i=1,2,...,B$, create SVS $\widehat{q_{\mathbf{H}_{S(\boldsymbol{\tau}_{i})}}}(\mathbf{x})$ and calculate its fitness value $V(\boldsymbol{\tau}_{i}^{(g-1)}, \mathbf{H}_{S(\boldsymbol{\tau}_{i})}^{(g-1)})$ along with the optimal bandwidth matrix $\mathbf{H}_{S(\boldsymbol{\tau}_{i})}^{(g-1)}$.
\item Sort the elements in $\mathbf{D}^{(g-1)} = \{\mathbf{D}_{1}^{(g-1)}, \mathbf{D}_{2}^{(g-1)}, ..., \mathbf{D}_{B}^{(g-1)} \}$ in descending order according to their fitness values $V(\boldsymbol{\tau}_{i}^{(g-1)}, \mathbf{H}_{S(\boldsymbol{\tau}_{i})}^{(g-1)})$, $i=1,2,...,B$, and rename the resulting sequence as $\mathbf{D}^{(g)} = \{\mathbf{D}_{1}^{(g)}, \mathbf{D}_{2}^{(g)}, ..., \mathbf{D}_{B}^{(g)} \}$.
\item Make the replica $\mathbf{D}^{+(g)} \equiv \mathbf{D}^{(g)}$.
\item Breed two new subsets using the pair of subsets $\mathbf{D}_{2k-1}^{(g)}$ and $\mathbf{D}_{2k}^{(g)}$, $k=1,2,...,B/2$; each pair of indices $\iota_{2k-1, j}^{(g)}$ and $\iota_{2k, j}^{(g)}$, $j = 1,2,...,b$, faces either of the following with a certain probability.
\begin{enumerate}[(i)]
\item {\it{Mutation}} : With mutation probability $p_{m}$, $\iota_{2k-1, j}^{(g)}$ and $\iota_{2k, j}^{(g)}$ are respectively replaced with the two indices randomly chosen from $\mathbf{D}_{1}^{+(g)} = \{\iota_{1, 1}^{+(g)}, \iota_{1, 2}^{+(g)}, ..., \iota_{1, b}^{+(g)} \}$.
\item {\it{Uniform crossover}} : $\iota_{2k-1, j}^{(g)}$ is swapped for $\iota_{2k, j}^{(g)}$ with crossover probability $p_{u}$.
\item {\it{Reproduction}} : $\iota_{2k-1, j}^{(g)}$ and $\iota_{2k, j}^{(g)}$ remain unchanged with probability $1-p_{u}-p_{m}$.
\end{enumerate}
\item For each renewed subset $\mathbf{D}_{i}^{(g)}$, for $i=1,2,...,B$, calculate the fitness value $V(\boldsymbol{\tau}_{i}^{(g)}, \mathbf{H}_{S(\boldsymbol{\tau}_{i})}^{(g)})$ along with ${\boldsymbol{\tau}_{i}^{(g)}}$ and the optimal bandwidth matrix $\mathbf{H}_{S(\boldsymbol{\tau}_{i})}^{(g)}$. Then, sort the renewed subsets in descending order by their renewed fitness values, and rename the resulting sequence as $\mathbf{D}^{*(g)} = \{\mathbf{D}_{1}^{*(g)}, \mathbf{D}_{2}^{*(g)}, ..., \mathbf{D}_{B}^{*(g)} \}$.
\item The renewed population $\mathbf{D}^{(g)} = \{\mathbf{D}_{1}^{+(g)}, \mathbf{D}_{2}^{+(g)}, ..., \mathbf{D}_{p_{e}B}^{+(g)}, \mathbf{D}_{1}^{*(g)}, \mathbf{D}_{2}^{*(g)}, ..., \mathbf{D}_{(1-p_{e}) B}^{*(g)} \}$ is carried over to generation $g$, where $p_{e}$ denotes the proportion of the $B$ subsets that are carried over to the next generation according to the elite selection rule.
\end{enumerate}
{\bf{Step~3:}} Completion of the algorithm at $g=G$:
\begin{enumerate}
\item Let us define $S^{*} \equiv S(\boldsymbol{\tau}^{*})$. Accept the SVS exhibiting the best fitness value $V(\boldsymbol{\tau}^{*}, \mathbf{H}_{S^{*}}^{*})$ at the final generation $G$ written as
\begin{eqnarray} \label{KDE.best.variables}
\widehat{q_{\mathbf{H}_{S^{*}}^{*}}}(\mathbf{x}) = \Biggl[ \frac{1}{N} \sum_{i=1}^{N} \prod_{j \in S^{*}} \frac{K_{h_{S^{*}}} \Bigl( x_{j} - X_{ij} \Bigr)^{\tau_{j}^{*}}}{C_{h_{S^{*}}}(\tau_{j}^{*})} \Biggr] \Biggl[ \prod_{j \notin S^{*}} \Biggl\{ \frac{1}{N} \sum_{i=1}^{N} K_{h_{j}} \Bigl( x_{j} - X_{ij} \Bigr) \Biggr\} \Biggr], \nonumber
\end{eqnarray}
along with the resulting subset $\mathbf{D}^{*} = \{\iota_{1}^{*}, \iota_{2}^{*}, ..., \iota_{b}^{*} \}$ and bandwidth matrix $\mathbf{H}_{S^{*}}^{*}$.
\end{enumerate}
\medskip
{\Remark{\hspace{-2.0mm}{\bf{.}}} \label{remark.ISE}} \hspace{-4.5mm} We modify the Least Square Cross-validation (LSCV) in Rudemo (\citeyear{Rudemo.1982}) and Bowman (\citeyear{Bowman.1984}) to our fitness function. Expanding ISE, we obtain
\begin{eqnarray} \label{def.ISE}
\lefteqn{\mathrm{ISE} \Bigl( \widehat{q_{\mathbf{H}_{S}}}(\cdot), f(\cdot) \Bigr)} \nonumber \\
&=& \int_{\mathbb{R}^{d}} \biggl[ \widehat{q_{\mathbf{H}_{S}}}(\mathbf{t}) - f(\mathbf{t}) \biggr]^2 d\mathbf{t} \nonumber \\
&=& \int_{\mathbb{R}^{d}} \widehat{q_{\mathbf{H}_{S}}}(\mathbf{t})^{2} d\mathbf{t} - 2 \int_{\mathbb{R}^{d}} \widehat{q_{\mathbf{H}_{S}}}(\mathbf{t}) f(\mathbf{t}) d\mathbf{t} + R(f).
\end{eqnarray}
Replacing the second term in \eqref{def.ISE} with its empirical form in a leave-one-out manner, and excluding the term $R(f)$ which is not involved in optimization, we obtain the fitness function
\begin{eqnarray} \label{def.CV}
\lefteqn{-V(\boldsymbol{\tau}, \mathbf{H}_{S})} \nonumber \\
&=& \int_{\mathbb{R}^{d}} \widehat{q_{\mathbf{H}_{S}}}(\mathbf{t})^{2} d\mathbf{t} \nonumber \\
& & - \frac{2}{N} \sum_{i_{0} = 1}^{N} \Biggl[ \frac{1}{N-1} \sum_{i_{1} = 1}^{N} I(i_{0} \neq i_{1})  \prod_{j \in S} \frac{K_{h_{S}} \Bigl( X_{i_{0}j} - X_{i_{1}j} \Bigr)^{\tau_{j}}}{C_{h_{S}}(\tau_{j})} \Biggr] \nonumber \\
& & \hspace{40mm} \times \prod_{j \notin S} \Biggl[  \frac{1}{N-1} \sum_{i_{2}=1}^{N} I(i_{0} \neq i_{2}) K_{h_{j}} \Bigl( X_{i_{0}j} - X_{i_{2}j} \Bigr) \Biggr], \nonumber \\
& & \mbox{subject to\ } \sum_{j \in S} \tau_{j} = b, \ \ \ \tau_{j} \in \{ 0, 1, 2, \ldots, b \}, \nonumber
\end{eqnarray}
where $I(\cdot)$ is an indicator function that takes the value one when the statement inside the braces is true.
\section{Numerical experiments} \label{numerical.experiments}

We perform the numerical experiments to validate our proposed GA. We consider two $d$-dimensional experimental cases, referred to as Type~J1 and Type~AR1. Type~J1 is a $d$-dimensional extension of Type~J from Wand and Jones (\citeyear{Wand.1993}), in which the marginal distribution of the first two variables coincides with Type~J, while the remaining $d-2$ variables are mutually independent and each follows a normal distribution. Therefore, the population-optimal active index set of Type~J1 is $S_{0} = \{1, 2 \}$, in which case $\mathcal{E}(f, q_{S_{0}}) = 0$. Type~AR1 is designed so that every pair among the $d$ normal variables exhibits correlation, with variance-covariance matrix $\boldsymbol{\Sigma}_{AR1, d}$ defined by a first-order autoregressive (AR(1)) structure, representing a setting in which full joint KDE minimizes the approximation error. It provides an interesting scenario in which, although the full active set minimizes the population approximation error, a smaller active index set with $|S| < d$ may yield a lower finite-sample estimation error because the reduction in variance can outweigh the increase in approximation error $\mathcal{E}(f,q_{S}) \neq 0$. Throughout the study, we employ Gaussian product kernel in (\ref{Gaussian.product.ker}). \\
\\{\bf{Type~J1:}}
\begin{eqnarray}
f_{J1, d}(\mathbf{x}) &=& \frac{1}{3}\, \phi_{d} \!\left(\mathbf{x}; \boldsymbol{\mu}_{J1, 1}, \boldsymbol{\Sigma}_{J1, 1} \right) + \frac{1}{3}\, \phi_{d} \!\left(\mathbf{x}; \boldsymbol{\mu}_{J1, 2}, \boldsymbol{\Sigma}_{J1, 2} \right) + \frac{1}{3}\, \phi_d \!\left(\mathbf{x}; \boldsymbol{\mu}_{J1, 3}, \boldsymbol{\Sigma}_{J1, 3} \right), \nonumber \\
\boldsymbol{\mu}_{J1, 1}^\top &=& \left(-\frac{6}{5},\,0,\,0,\ldots,0 \right), \boldsymbol{\mu}_{J1, 2}^\top = \left(\frac{6}{5},\,0,\,0,\ldots,0 \right), \boldsymbol{\mu}_{J1, 3}^\top = \left(0,\,0,\,0,\ldots,0 \right), \nonumber \\
\boldsymbol{\Sigma}_{J1, k}
&=& {\sigma}^{2}
\begin{pmatrix}
\mathbf{J1}_{k} & \mathbf{0} \\
\mathbf{0} & \mathbf{I}_{d-2}
\end{pmatrix}, \ 
\mathbf{J1}_{k} =
\begin{pmatrix}
1 & \rho_k \\
\rho_k & 1
\end{pmatrix},\ 
\sigma = \frac{3}{5},\ \rho_{1} = \frac{7}{10},\ \rho_{2} = \frac{7}{10},\ \rho_{3} = -\frac{7}{10}. \nonumber
\end{eqnarray}
\\{\bf{Type~AR1:}}
\begin{eqnarray}
f_{\mathrm{AR1},d}(\mathbf{x}) &=& \phi_d(\mathbf{x}; \mathbf{0}_d, \boldsymbol{\Sigma}_{\mathrm{AR1},d}), \nonumber \\
\left(\mathbf{\Sigma}_{\mathrm{AR1},d}\right)_{ij} &=& \rho^{|i-j|}, \ \ \ i,j=1,\ldots,d, \ \ \ \rho = \frac{2}{5}. \nonumber
\end{eqnarray}

We numerically computed the oracle-calibrated MISE values for SVS and the full-joint KDE under Type~J1 and AR1 with $d=10$, as reported in \hyperref[tab:typeJ1_calibration]{Tables~\ref*{tab:typeJ1_calibration}} and \hyperref[AR1_d10_all_combinations_subset]{~\ref*{AR1_d10_all_combinations_subset}} respectively. Since both settings are based on Gaussian distributions, the MISE can be evaluated analytically via Gaussian convolution. For Type~J1, the finite-sample MISE is evaluated exactly. For Type~AR1, we use a calibrated finite-sample MISE in which the dependence among the component KDEs arising from their construction from the same sample is neglected. The bandwidths are selected by numerically minimizing the corresponding MISE criterion. In the AR1 setting, the subset $S$ is selected by an exhaustive search over all possible subsets, and the subset yielding the smallest calibrated MISE, denoted by $S_{\mathrm{MISE}}^{*}$, is adopted.
\begin{table}[htbp]
\centering
\begin{tabular}{llllll}
\hline
\hline
$N$ & $h_{S}^{*}$ & $h_{m}^{*}$ & SVS & Full-joint & Ratio \\
\hline
100  & 0.2727 & 0.2237 & 0.836\ \ \ \ & $2.442$\ \ \ \ & 0.343 \\
250  & 0.2238 & 0.1768 & 0.499 & 2.189 & 0.228 \\
500  & 0.1945 & 0.1489 & 0.328 & 1.994 & 0.164 \\
1000\ \ \ \ \ & 0.1700\ \ \ \ \ & 0.1261\ \ \ \ \ & 0.212 & 1.801 & 0.118 \\
10000\ \ \ \ \ & 0.1119\ \ \ \ \ & 0.0746\ \ \ \ \ & 0.045 & 1.216 & 0.037 \\
100000\ \ \ \ \ & 0.0753\ \ \ \ \ & 0.0452\ \ \ \ \ & 0.009 & 0.766 & 0.012 \\
\hline
\hline
\end{tabular}
\caption{Oracle-calibrated MISE values ($\times 10^{-4}$) for SVS and the full-joint KDE under Type~J1 with $d=10$, using the population-optimal subset $S_{0} = \{1,2\}$ and the corresponding optimal bandwidths $h_{S}^{*}$ and $h_{m}^{*}$. The ratio is defined by $\mbox{MISE(SVS)/MISE(full)}$.}
\label{tab:typeJ1_calibration}
\end{table}
\begin{table}[tb]
\centering
\begin{tabular}{l l l l l l l l}
\hline
\hline
$N$ & $|S_{\mathrm{MISE}}^{*}|$ & $S_{\mathrm{MISE}}^{*}$ & $h_{S}^{*}$ & $h_{m}^{*}$ & SVS & Full-joint & Ratio \\
\hline
100      & 3  & $\{8,9,10\}$               & 0.4717 & 0.3070 & 3.418 & 5.080 & 0.673 \\
250      & 3  & $\{8,9,10\}$               & 0.3954 & 0.2295 & 2.913 & 4.537& 0.642 \\
500      & 4  & $\{7,8,9,10\}$             & 0.4045 & 0.1795 & 2.633 & 4.121 & 0.639 \\
1000     & 4  & $\{7,8,9,10\}$           & 0.3622 & 0.1442 & 2.400  & 3.710 & 0.647 \\
10000    & 6  & $\{5,6,7,8,9,10\}$     & 0.3444 & 0.0658 & 1.853  & 2.477 & 0.748 \\
1000000  & 10 & $\{1,2,3,4,5,6,7,8,9,10\}$ & 0.3288 & - & 0.915 & 0.915 & 1.000 \\
\hline
\hline
\end{tabular}
\caption{Oracle-calibrated MISE values ($\times 10^{-6}$) over all subsets $S$ under the AR(1)-type Gaussian model with $d=10$ and $\Sigma_{ij}=0.4^{|i-j|}$. The ratio is defined by $\mbox{MISE(SVS)/MISE(full)}$.} \label{AR1_d10_all_combinations_subset}
\end{table}

In Type~J1, the oracle calibration results suggest that SVS outperforms the full joint KDE in terms of MISE for all values of $N$. In Type~AR1 setting, the oracle calibration results also suggest that the SVS estimator can outperform the full joint KDE in terms of MISE for small sample sizes, while its relative advantage is expected to diminish for sufficiently large $N$. This behavior can be explained by the bias-variance trade-off: for small sample sizes, the reduction in variance achieved by SVS outweighs the increase in approximation error, whereas for large sample sizes the approximation error becomes relatively more influential as the variance decreases.

Subsequently, we conduct Monte-Carlo simulation. In the simulation, we generate samples of size $N=250$ and $500$ with $d=10$, and for each generated dataset, construct the SVS ten times and calculate the average of the resulting 10 estimators at every generation $g$. The respective averages are denoted by $\mathrm{LSCV}^{*}(g)$ and $\mathrm{ISE}^{*}(g)$. We set the GA parameters as follows: $B=50$, $G=100$, $p_{u}=0.5$, $p_{m}=0.05$, and $p_{elite}=0.1$. These parameters are tuned with reference to parameter values that have been empirically used in previous studies of GA. We conduct experiments for $b=5$ and $10$. In Type~J1, we also calculate the correct selection rate (CSR) with its S.D. at every $g$, defined as the proportion of the total sharpness allocated to the variables in the population-optimal active set $S_{0} = \{ 1, 2\}$: $\sum_{j \in S_{0}}\tau_{j}/b$.

The results of Type~J1 are given in \hyperref[Monte.Carlo.J1.results]{Table~\ref*{Monte.Carlo.J1.results}}, together with the ISE values of the full-joint KDE and PPDE computed using the same sample as that used for SVS. For PPDE, the iterative procedure converged after three iterations, although the maximum number of iterations is set to 50. The panels (a) and (b), (c) and (d), and (e) and (f) of \hyperref[fig.results.Type.J1]{Figure~\ref*{fig.results.Type.J1}} show the $\mathrm{LSCV}^{*}$, $\mathrm{ISE}^{*}$, and $\mathrm{CSR}$ values, respectively, over generations. The panels on the left and right correspond to the cases of $N=250$ and $N=500$, respectively. The $\mathrm{LSCV}^{*}$ and $\mathrm{ISE}^{*}$ values are shown for the representative case of $b=5$, whereas the $\mathrm{CSR}$ values are shown for $b=4,\ldots,10$. The simulation results show that SVS can outperform both the full-joint KDE and PPDE in terms of $\mathrm{ISE}^{*}$. These panels indicate that the sharpening parameters are consistently concentrated on the population-optimal active set $S_{0}$ when $b$ is not too large. 
\begin{table}
\begin{center}
{\scriptsize{
\begin{tabular}{llllllllll}
\hline
\hline
$g$ & 1 & 25 & 50 & 75 & 100 & Full-joint &  PPDE \\
\hline
$\underline{N=250}$ & --- & --- & --- & --- &  --- & 2199 (---) & 1069 (---) \\
$b=5$ & --- & --- & --- & --- & --- \\
$\mbox{LSCV}^{*}$ & -2561 (109) & -2739 (0) & -2739 (0) & -2739 (0) & -2739 (0) \\
$\mathrm{ISE}^{*}$ & \phantom{-}921 (189) & \phantom{-}572 (0) & \phantom{-}572 (0) & \phantom{-}572 (0) & \phantom{-}572 (0) \\
$\mbox{CSR}$ & \phantom{-}0.480 (.286) & \phantom{-}1.000 (0) & \phantom{-}1.000 (0) & \phantom{-}1.000 (0) & \phantom{-}1.000 (0) \\
$b=10$ & --- & --- & --- & --- & --- \\
$\mbox{LSCV}^{*}$ & -2193  (56) & -2712 (26) & -2715 (21) & -2720 (12) & -2720 (12) \\
$\mathrm{ISE}^{*}$ & \phantom{-}1230 (90) & \phantom{-}696 (125) & \phantom{-}689 (107) & \phantom{-}668 (70) & \phantom{-}668 (70) \\
$\mbox{CSR}$ & \phantom{-}0.310 (.129) & \phantom{-}0.710 (.228) & \phantom{-}0.700 (.216) & \phantom{-}0.720 (.193) & \phantom{-}0.720 (.193) \\
\hline
$\underline{N=500}$ & --- & --- & --- & --- & --- & 1967 (---) & 900 (---) \\
$b=5$ & --- & --- & --- & --- & --- \\
$\mbox{LSCV}^{*}$ & -2729 (49) & -2855 (0) & -2855 (0) & -2855 (0) & -2855 (0) \\
$\mathrm{ISE}^{*}$ & \phantom{-}611 (77) & \phantom{-}378 (0) & \phantom{-}378 (0) & \phantom{-}378 (0) & \phantom{-}378 (0) \\
$\mbox{CSR}$ & \phantom{-}0.580 (.140) & \phantom{-}1.000 (0) & \phantom{-}1.000 (0) & \phantom{-}1.000 (0) & \phantom{-}1.000 (0) \\
$b=10$ & --- & --- & --- & --- & --- \\
$\mbox{LSCV}^{*}$ & -2463 (70) & -2838 (20) & -2847 (14) & -2852 (11) & -2852 (11) \\
$\mathrm{ISE}^{*}$ & \phantom{-}947 (110) & \phantom{-}433 (59) & \phantom{-}422 (60) & \phantom{-}399 (49) & \phantom{-}399 (49) \\
$\mbox{CSR}$ & \phantom{-}0.410 (.099) & \phantom{-}0.860 (.151) & \phantom{-}0.880 (.155) & \phantom{-}0.940 (.126) & \phantom{-}0.940 (.126) \\
\hline
\hline
\end{tabular}}}
\caption{Results of numerical experiments Type~J1: Standard deviations are shown in parentheses. For the full-joint KDE and PPDE, only the $\mathrm{ISE}$ values are reported. All values other than $\mathrm{CSR}$ are expressed in units of $10^{-7}$.}  \label{Monte.Carlo.J1.results}
\end{center}
\end{table}
\begin{figure}[h]
\begin{center}
    \begin{minipage}[t]{0.38\hsize}
        \center
        \captionsetup{width=.95\linewidth}
        \includegraphics[width=\textwidth]{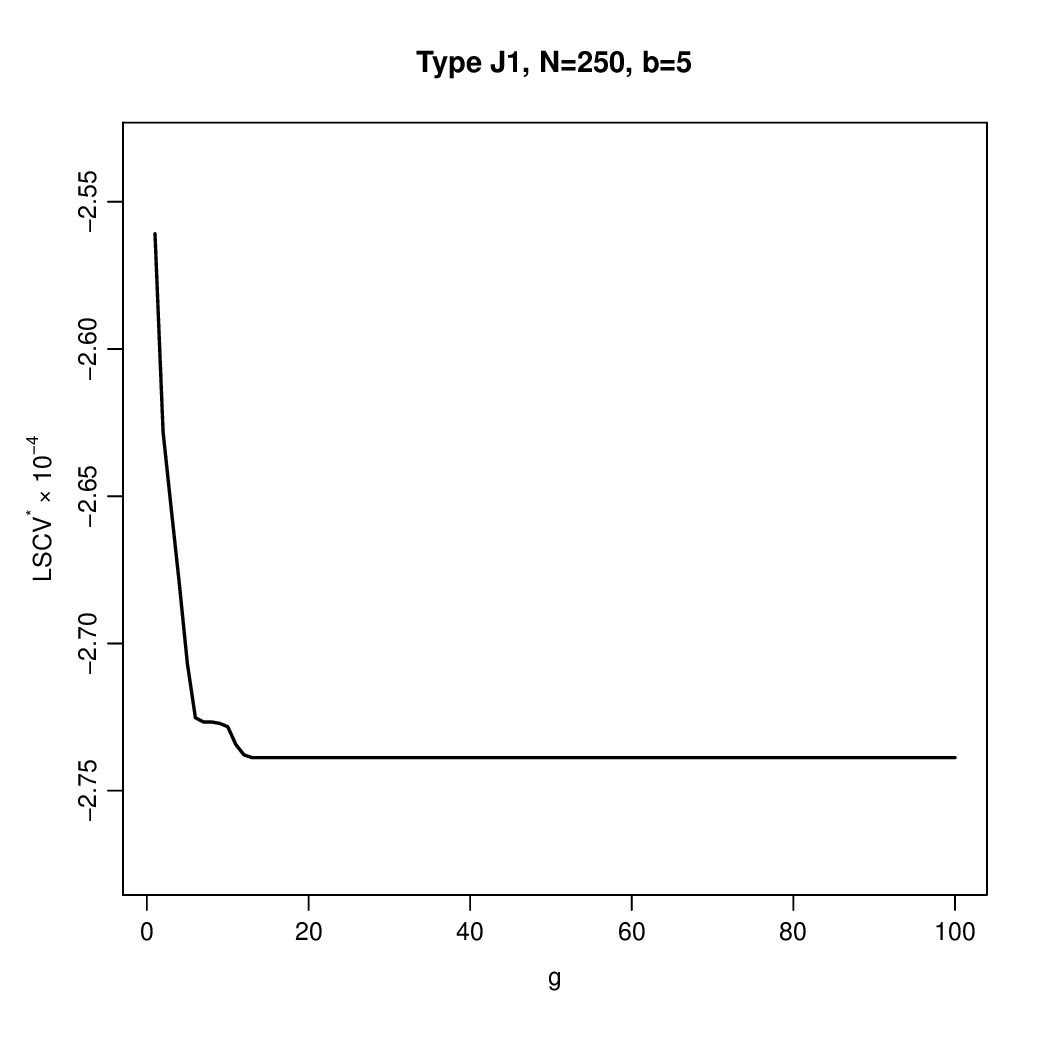}
        \mbox{(a)}
    \end{minipage}
    \begin{minipage}[t]{0.38\hsize}
        \center
        \captionsetup{width=.95\linewidth}
        \includegraphics[width=\textwidth]{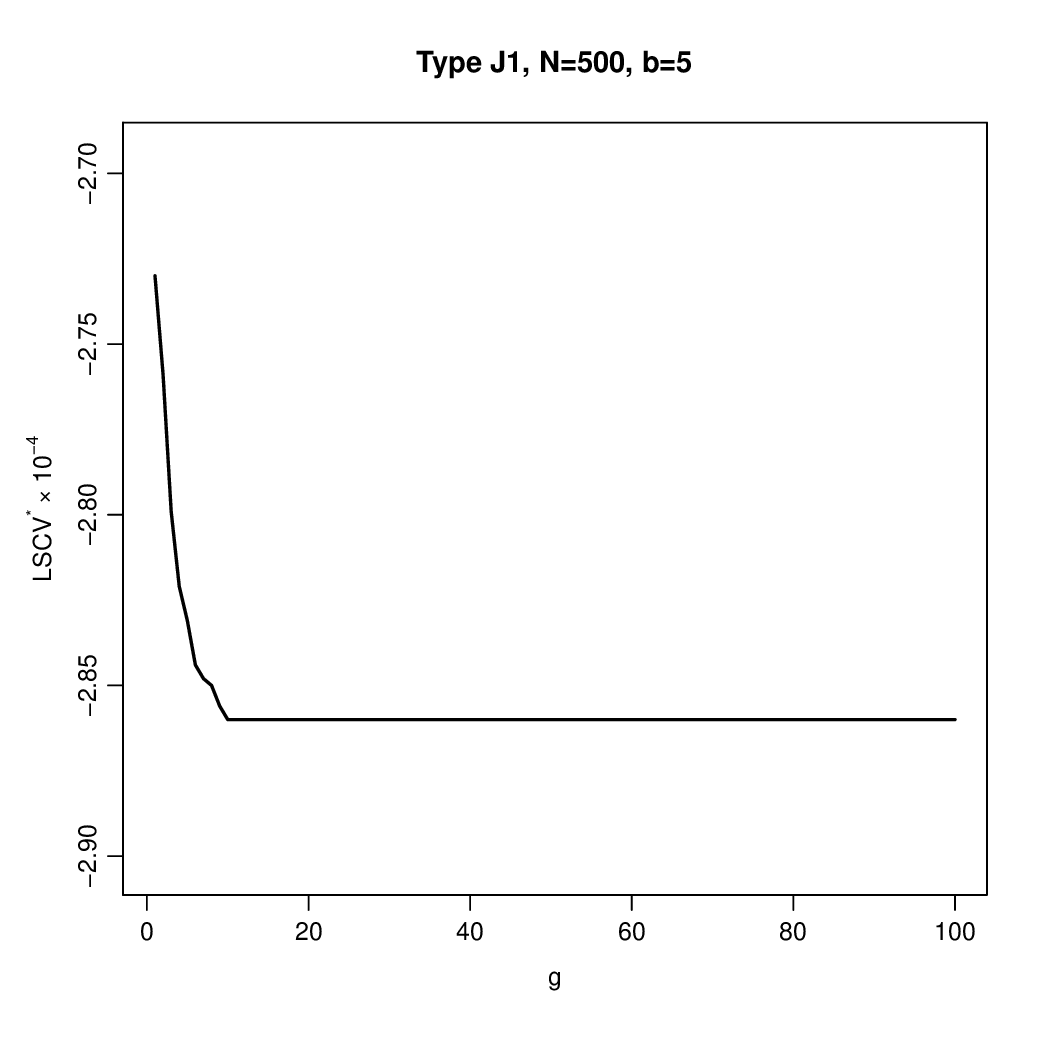}
        \mbox{(b)}
    \end{minipage}
    \begin{minipage}[t]{0.38\hsize}
        \center
        \captionsetup{width=.95\linewidth}
	\includegraphics[width=\textwidth]{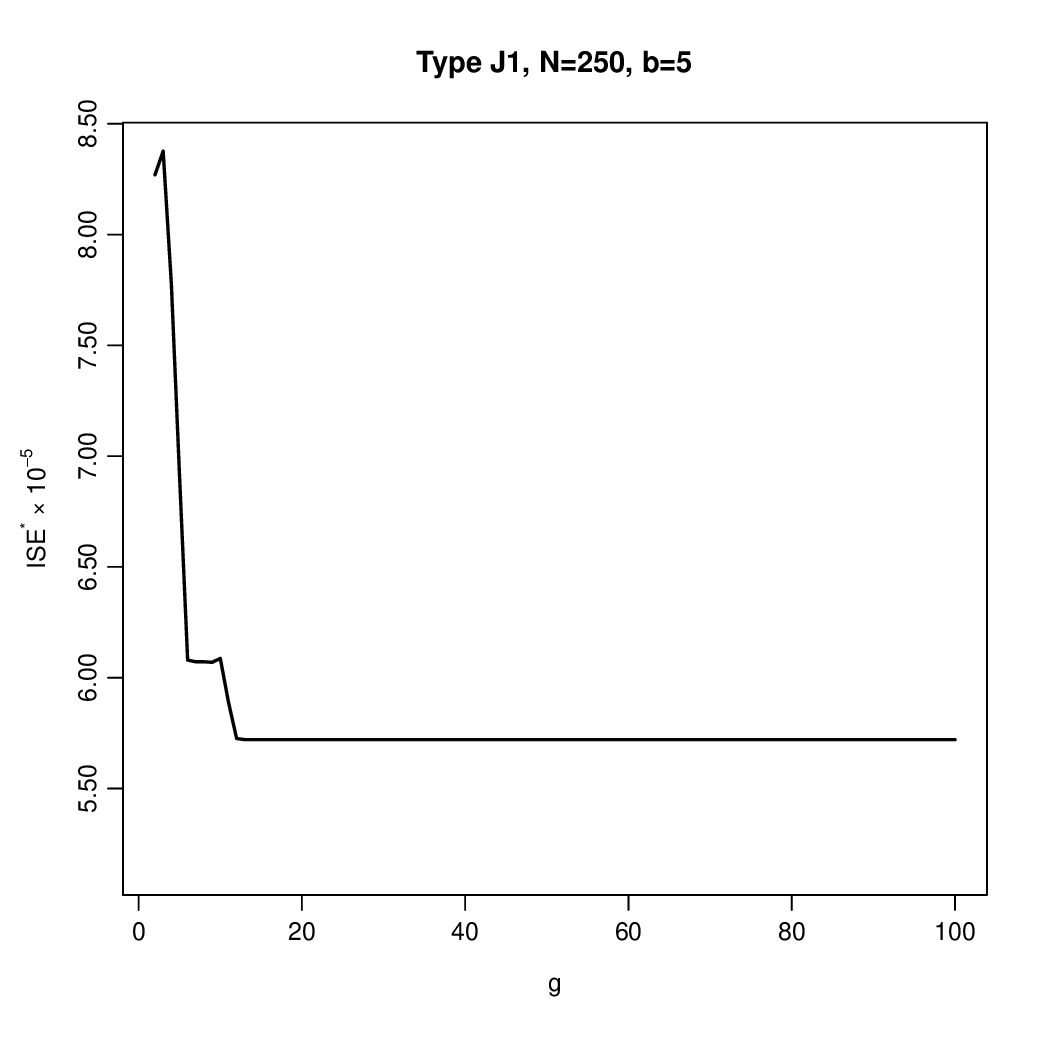}
       	\mbox{(c)}
    \end{minipage}
    \begin{minipage}[t]{0.38\hsize}
        \center
        \captionsetup{width=.95\linewidth}
        \includegraphics[width=\textwidth]{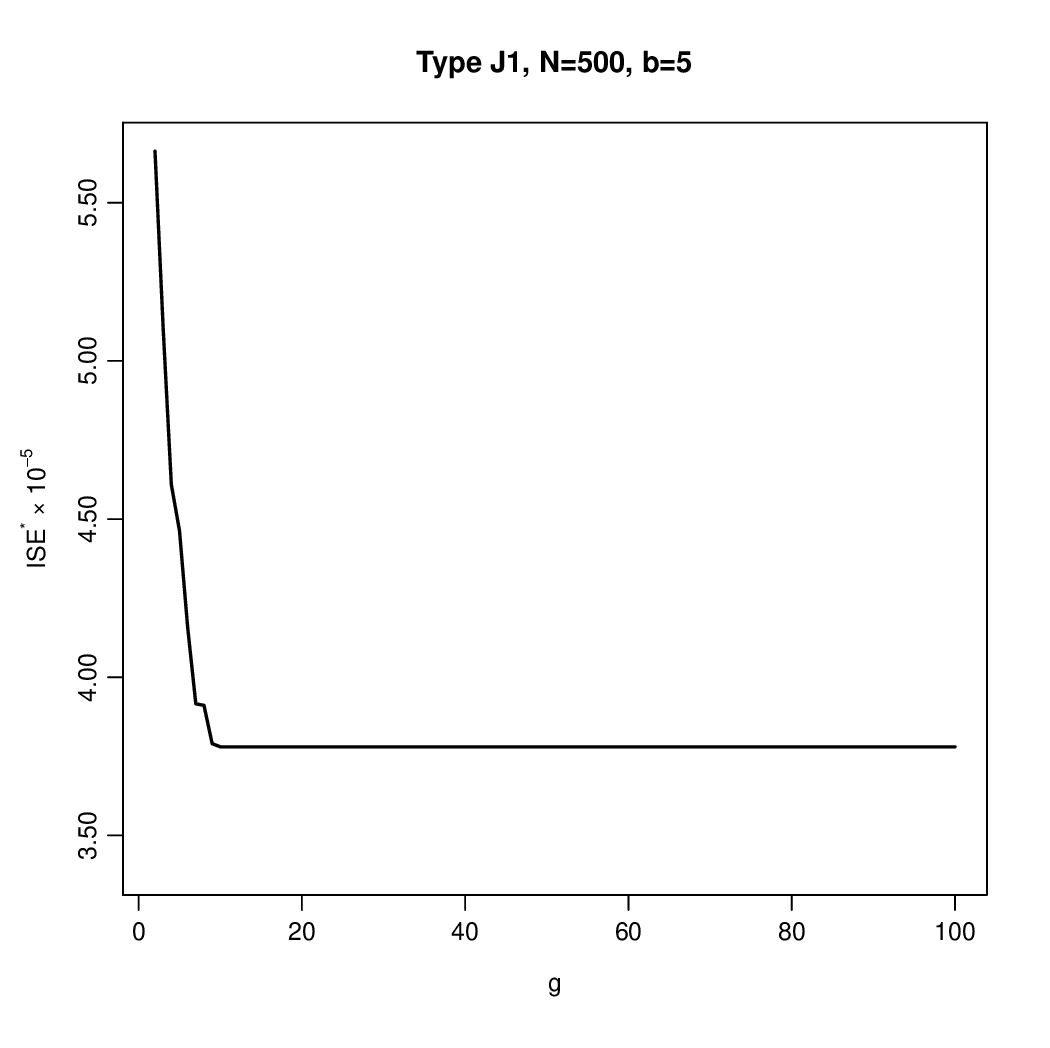}
      	\mbox{(d)}
    \end{minipage}
    \begin{minipage}[t]{0.38\hsize}
        \center
        \captionsetup{width=.95\linewidth}
        \includegraphics[width=\textwidth]{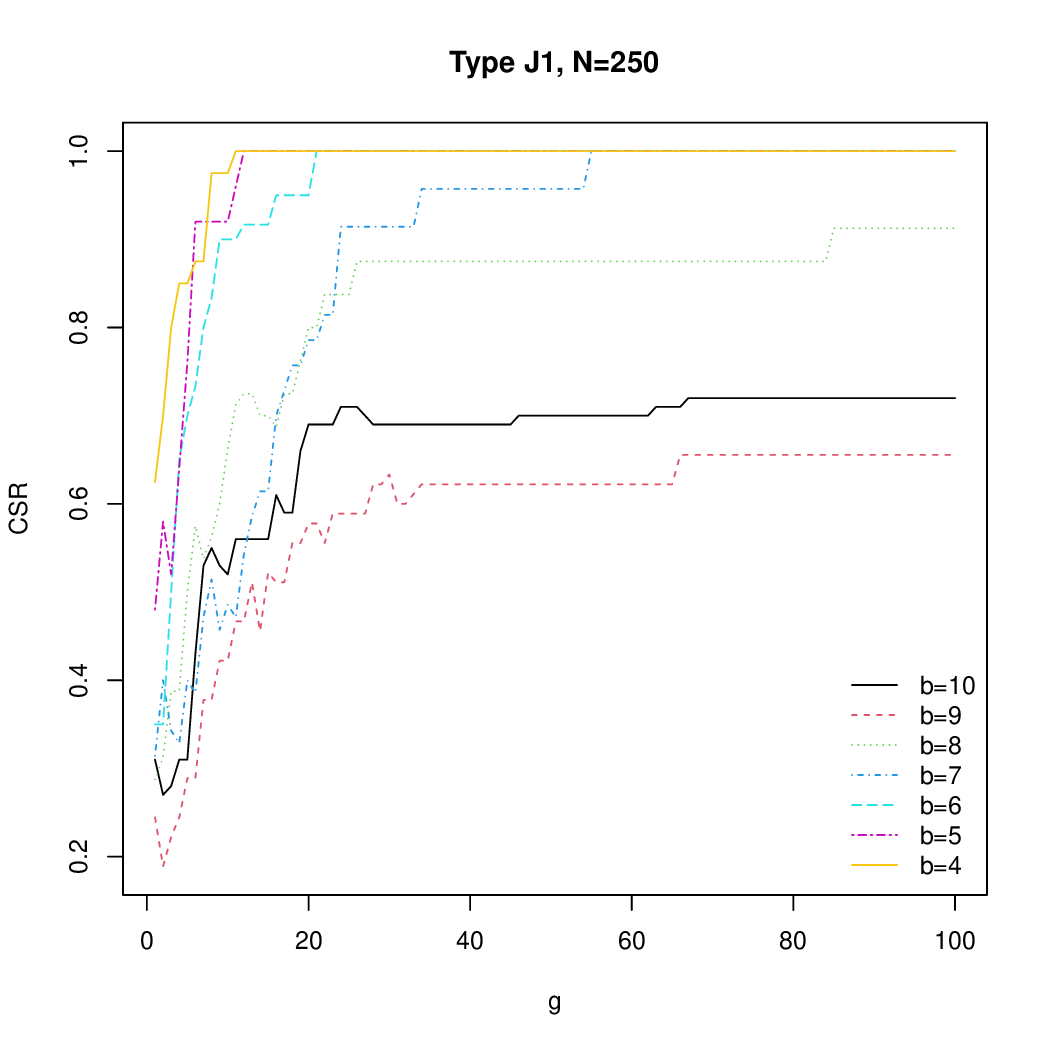}
        \mbox{(e)}
    \end{minipage}
    \begin{minipage}[t]{0.38\hsize}
        \center
        \captionsetup{width=.95\linewidth}
        \includegraphics[width=\textwidth]{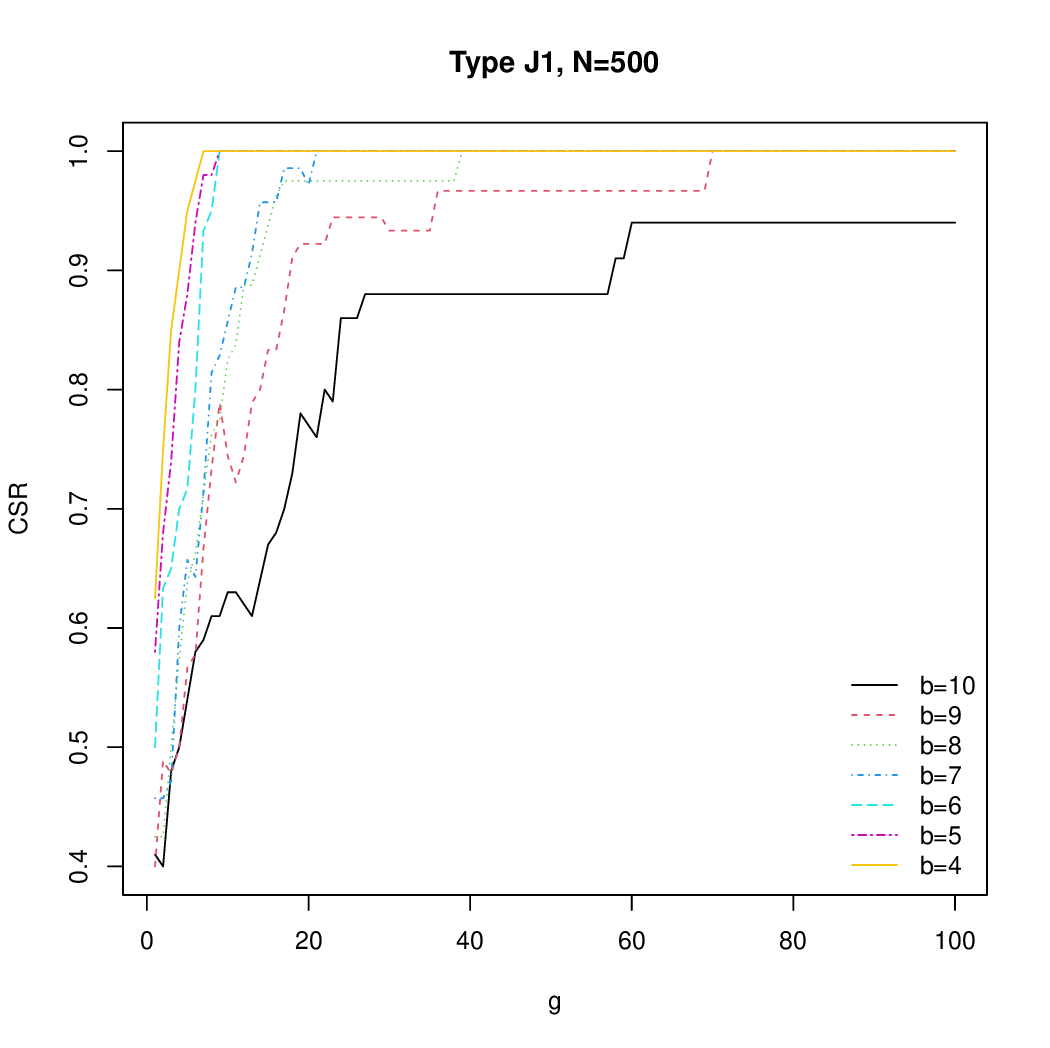}
        \mbox{(f)}
    \end{minipage}
\caption{The results of simulation : Type~J1.} \label{fig.results.Type.J1}
\end{center}
\end{figure}

The results of Type~AR1 are given in \hyperref[Monte.Carlo.AR1.results]{Table~\ref*{Monte.Carlo.AR1.results}}, together with the ISE values of the full-joint KDE and PPDE. \hyperref[fig.results.Type.AR1]{Figure~\ref*{fig.results.Type.AR1}} is organized in the same manner as \hyperref[fig.results.Type.J1]{Figure~\ref*{fig.results.Type.J1}}, except that panels (e) and (f) present stacked area charts of the mean sharpening parameters $\tau_{i}$ across generations, averaged over 10 independent GA runs. At $g=100$, the ratios of the sharpening parameters are $\tau_{4} : \tau_{5} : \tau_{6} : \tau_{7} \approx 1 : 2 : 1 : 1$ for $N=250, b=5$, and $\tau_{1} : \tau_{2} : \tau_{4} : \tau_{5} : \tau_{6} : \tau_{8} : \tau_{9} : \tau_{10} \approx 1 : 1 : 4 : 4 : 12 : 1 : 1 : 1$ for $N=500, b=5$. The behavior of the estimated sharpness parameters is also consistent with the oracle calibration. As $N$ increases, the sharpness tends to be distributed across a wider range of variables rather than being concentrated on a small subset, suggesting a gradual shift toward a less sparse representation.

When $N=250$, SVS substantially outperforms the full-joint KDE in terms of ISE for both values of $b$. This improvement persists when the sample size increases to $N=500$. In particular, the final ISE of SVS decreases from $3.728 \times 10^{-6}$ to $3.025 \times 10^{-6}$ for $b=5$, and from $3.867 \times 10^{-6}$ to $3.211 \times 10^{-6}$ for $b=10$, whereas the corresponding ISEs of the full-joint KDE are $15.730 \times 10^{-6}$ and $8.795 \times 10^{-6}$, respectively. Thus, SVS consistently reduces the estimation error relative to the full-joint KDE under the AR1 setting, indicating that the variance reduction achieved through the sparse representation can remain beneficial even when the sparse approximation does not fully capture the true density and hence incurs a nonzero approximation error. Compared with PPDE, SVS does not outperform it in terms of ISE for either value of $N$. This is expected because Type~AR1 is essentially a Gaussian dependence structure, which is particularly suitable for PPDE. 
\begin{table}
\begin{center}
{\scriptsize{
\begin{tabular}{llllllllll}
\hline
\hline
$g$ & 1 & 25 & 50 & 75 & 100 & Full-joint &  PPDE \\
\hline
$\underline{N=250}$ & --- & --- & --- & --- &  --- & 15730 (---) & 979 (0) \\
$b=5$ & --- & --- & --- & --- & --- \\
$\mbox{LSCV}^{*}$ & -3218 (96) & -3396 (0) & -3396 (0) & -3396 (0) & -3396 (0) \\
$\mathrm{ISE}^{*}$ & \phantom{-}3742 (199) & \phantom{-}3728 (0) & \phantom{-}3728 (0) & \phantom{-}3728 (0) & \phantom{-}3728 (0) \\
$b=10$ & --- & --- & --- & --- & --- \\
$\mbox{LSCV}^{*}$ & -2921 (95) & -3506 (14) & -3516 (0) & -3516 (0) & -3516 (0) \\
$\mathrm{ISE}^{*}$ & \phantom{-}3983 (197) & \phantom{-}3989 (213) & \phantom{-}3867 (0) & \phantom{-}3867 (0) & \phantom{-}3867 (0) \\
\hline
$\underline{N=500}$ & --- & --- & --- & --- & --- & 8795 (---) & 359 (0) \\
$b=5$ & --- & --- & --- & --- & --- \\
$\mbox{LSCV}^{*}$ & -3983 (104) & -4100 (6) & -4100 (6) & -4100 (6) & -4100 (6) \\
$\mathrm{ISE}^{*}$ & \phantom{-}3104 (107) & \phantom{-}3025 (81) & \phantom{-}3025 (81) & \phantom{-}3025 (81) & \phantom{-}3025 (81) \\
$b=10$ & --- & --- & --- & --- & --- \\
$\mbox{LSCV}^{*}$ & -3734 (134) & -4085 (47) & -4098 (43) & -4110 (30) & -4119 (27) \\
$\mathrm{ISE}^{*}$ & \phantom{-}3238 (138) & \phantom{-}3152 (178) & \phantom{-}3189 (177) & \phantom{-}3196 (183) & \phantom{-}3211 (197) \\
\hline
\hline
\end{tabular}}}
\caption{Results of numerical experiments Type~AR1: Standard deviations are shown in parentheses. For the full-joint KDE and PPDE, only the ISE values are reported. All values are expressed in units of $10^{-9}$.}  \label{Monte.Carlo.AR1.results}
\end{center}
\end{table}
\begin{figure}[h]
\begin{center}
    \begin{minipage}[t]{0.38\hsize}
        \center
        \captionsetup{width=.95\linewidth}
        \includegraphics[width=\textwidth]{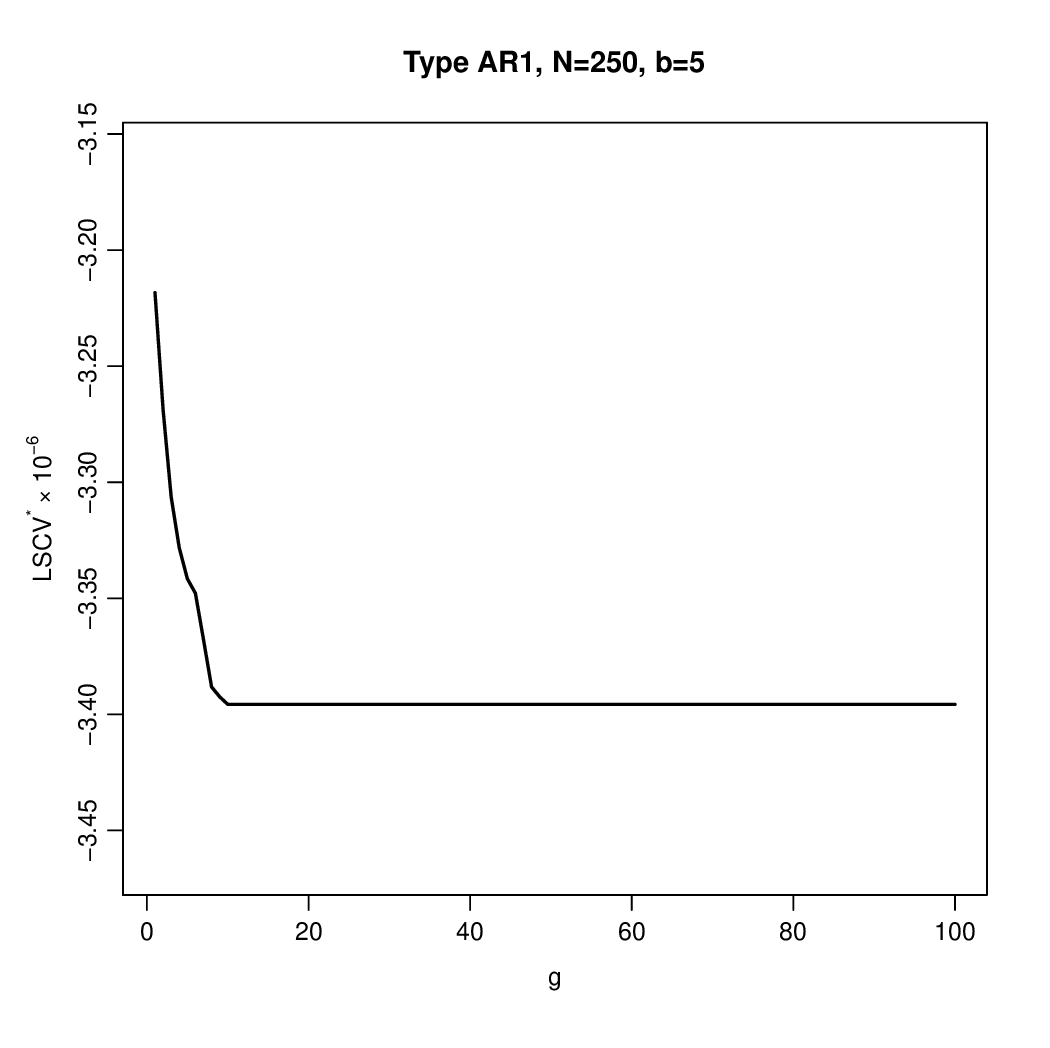}
      	\mbox{(a)}
    \end{minipage}
    \begin{minipage}[t]{0.38\hsize}
        \center
        \captionsetup{width=.95\linewidth}
        \includegraphics[width=\textwidth]{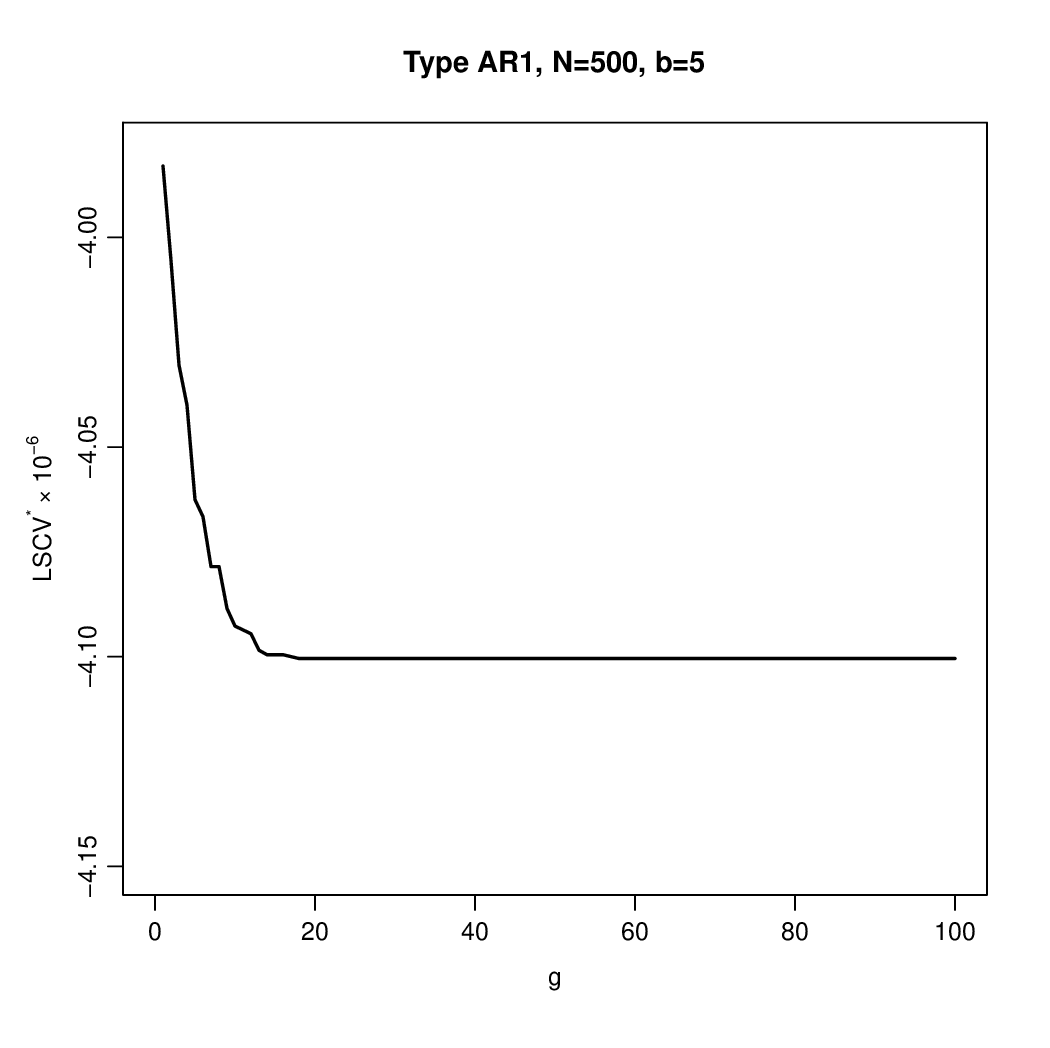}
      	\mbox{(b)}
    \end{minipage}
    \begin{minipage}[t]{0.38\hsize}
        \center
        \captionsetup{width=.95\linewidth}
        \includegraphics[width=\textwidth]{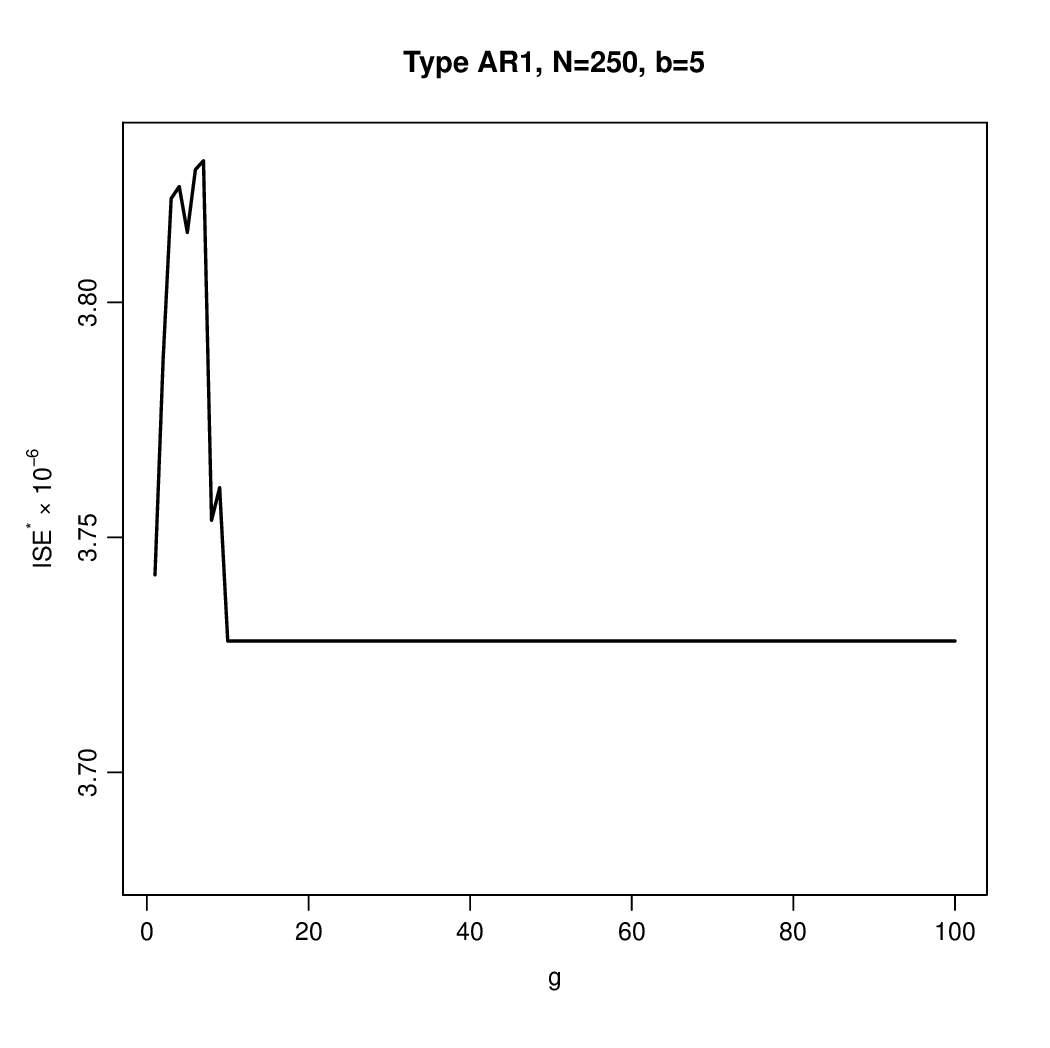}
      	\mbox{(c)}
    \end{minipage}
    \begin{minipage}[t]{0.38\hsize}
        \center
        \captionsetup{width=.95\linewidth}
        \includegraphics[width=\textwidth]{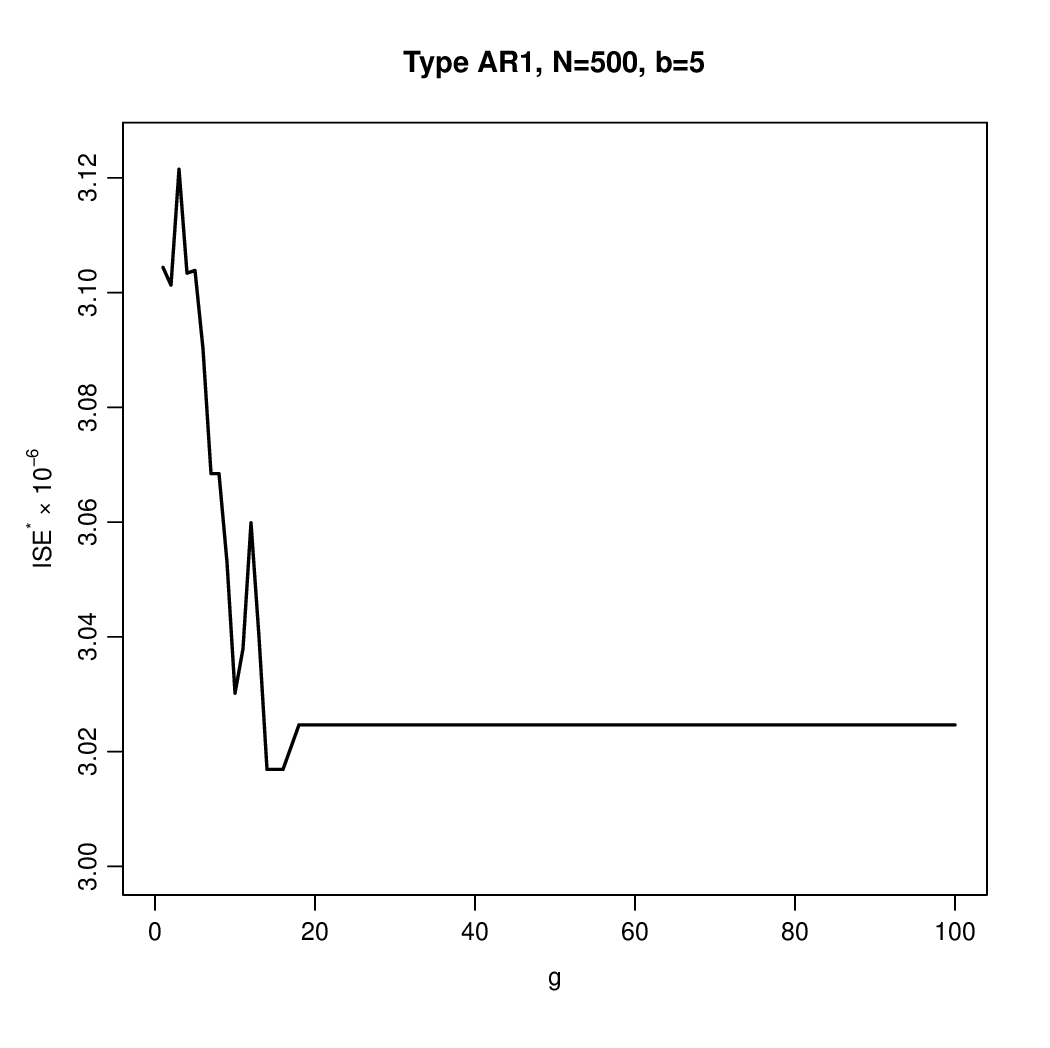}
      	\mbox{(d)}
    \end{minipage}
    \begin{minipage}[t]{0.38\hsize}
        \center
        \captionsetup{width=.95\linewidth}
        \includegraphics[width=\textwidth]{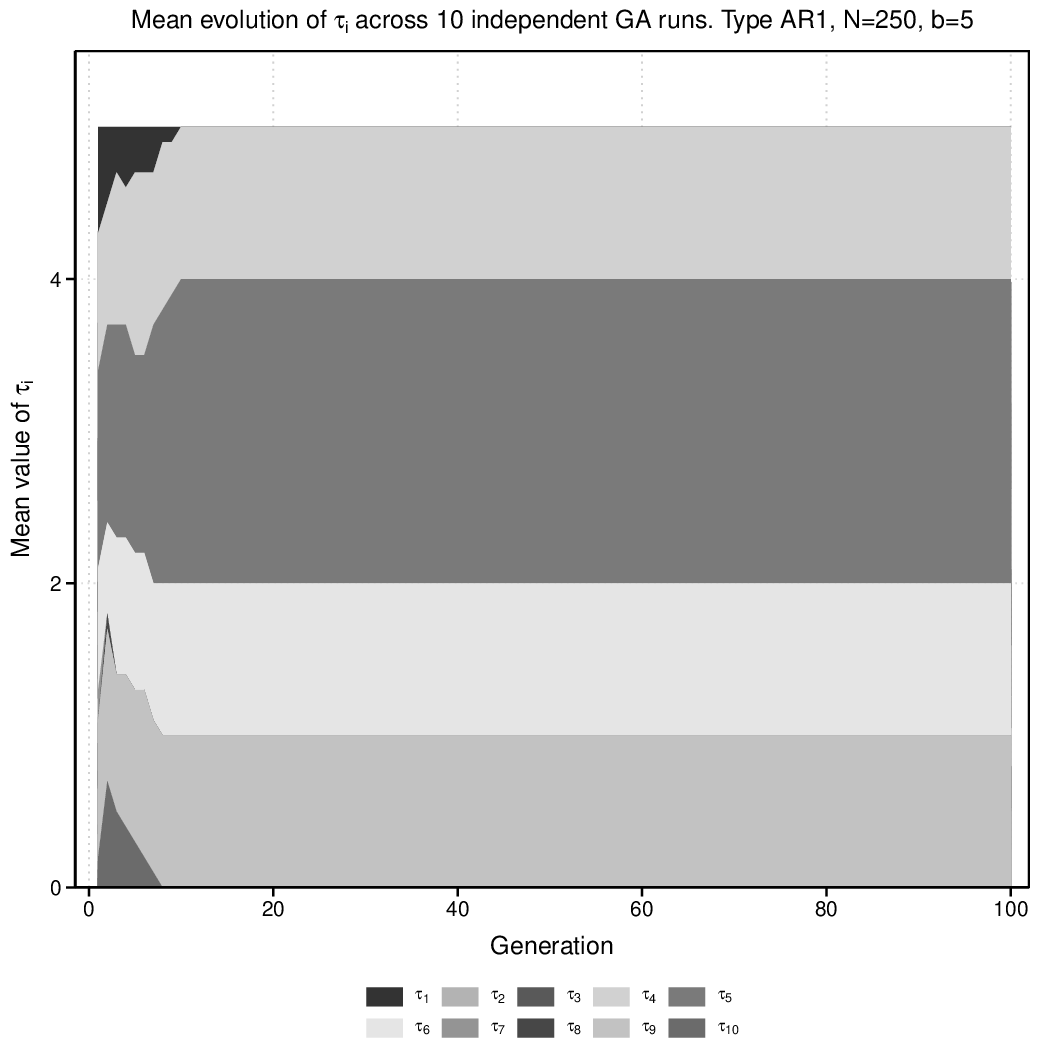}
      	\mbox{(e)}
    \end{minipage}
    \begin{minipage}[t]{0.38\hsize}
        \center
        \captionsetup{width=.95\linewidth}
        \includegraphics[width=\textwidth]{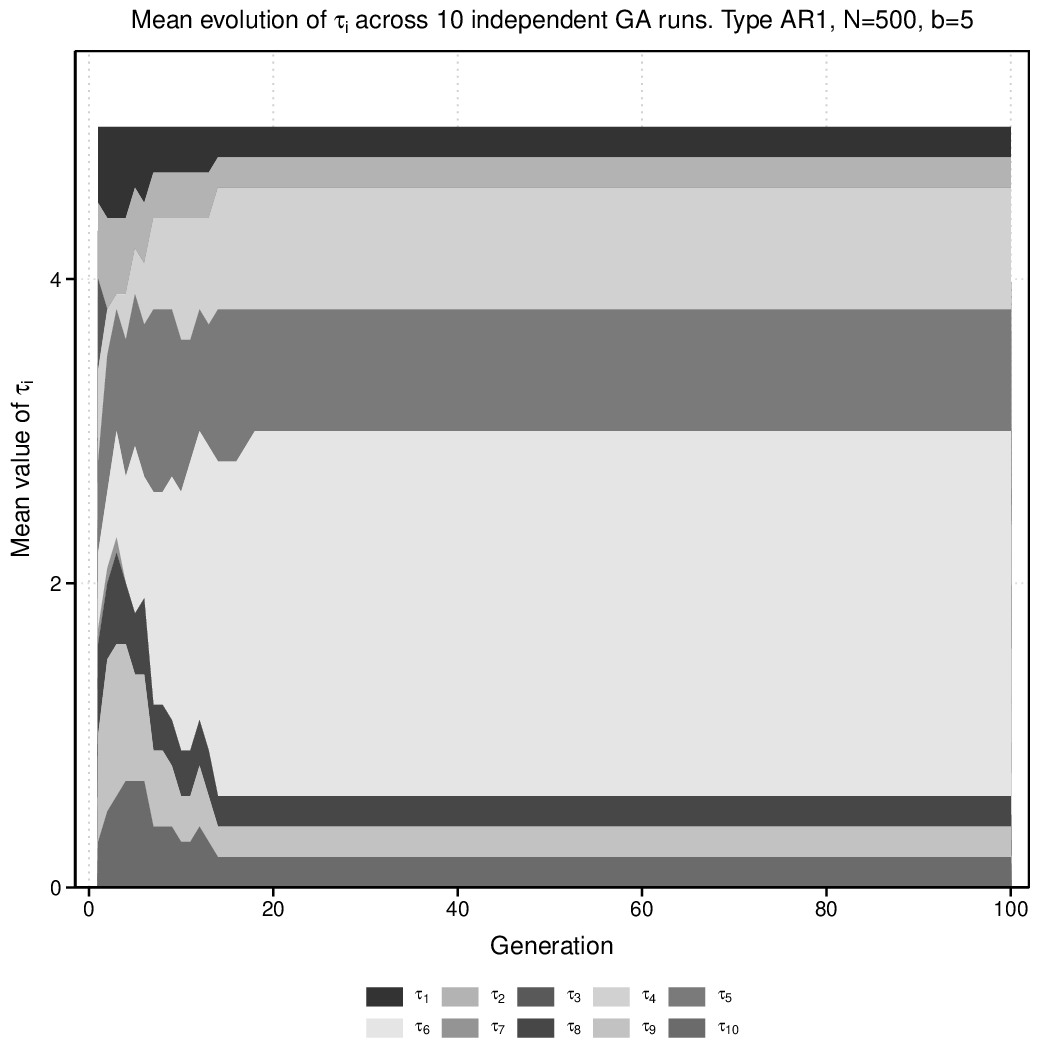}
      	\mbox{(f)}
    \end{minipage}
\caption{The results of simulation : Type~AR1.} \label{fig.results.Type.AR1}
\end{center}
\end{figure}
\section{Real data application} \label{real.data}

We show a real data application of our method. We use decathlon dataset, originally compiled by Chen and Samworth (\citeyear{Chen.2016}), which is distributed as the $\it{decathlon\_raw}$ dataset in the R package $\texttt{scar}$. The dataset consists of records for 614 male decathletes, each with measurements for the following 10 events: 100 metres (s), Long Jump (m), Shot Put (m), High Jump (m), 400 metres (s), 110 metres Hurdles (s), Discus Throw (m), Pole Vault (m), Javelin Throw (m) and 1500 metres (s). The variables are indexed according to the order of the events listed above. We apply a ${\it{whitening\ transformation}}$ (e.g., Kessy et al., \citeyear{Kessy.2018}) to the dataset before applying SVS to place all variables on a common scale. In addition, the whitening transformation removes linear correlations among the variables, allowing SVS to identify informative subsets of variables by exploiting the nonlinear dependencies that remain after whitening. In the estimation, we set $(b, B, G, p_{u}, p_{m}, p_{e}) = (10, 50, 100, 0.475, 0.05, 0.1)$ and implement our GA 10 times on the whitened decathlon dataset, reporting the average over the 10 runs.

The numerical results of applying SVS to the decathlon dataset are presented in \hyperref[tab.deca.results]{Table~\ref*{tab.deca.results}}, and \hyperref[fig.results.deca]{Figure~\ref*{fig.results.deca}} provides a graphical summary of the results. The panels (a) and (b) of \hyperref[fig.results.deca]{Figure~\ref*{fig.results.deca}} show the evolution of the LSCV criterion and the bandwidth for the joint part, respectively. The curves are averaged over 10 runs, and the gray shaded area represents $\pm$S.D.. The panel~(c) shows the mean values of $\tau_{i}$ over 10 independent GA runs at the final generation. The panel~(d) shows the evolution of the mean number of active $\tau_{i}$'s across 10 independent GA runs. Panel~(e) is a stacked area chart showing the mean values of the sharpening parameters $\tau_{i}$ across generations, averaged over 10 independent GA runs. Panel~(f) shows the final composition of $\tau_{i}$ for each of the 10 independent GA runs.

From the results, we notice SVS consistently identifies the same variable subset $\{\tau_{2},\tau_{8},\tau_{10} \}$, corresponding to $\{\mbox{Long Jump}, \mbox{Pole Vault}, \mbox{1500\ metres} \}$, as a locally dependent variable subset. The corresponding ratio is $\tau_{2} : \tau_{8} : \tau_{10} \approx 4.9 : 4.0 : 1.1$. It suggests that the extracted local variable structure is stable. Since linear correlations are removed by the whitening transformation, this result suggests that these variables retain nonlinear dependency structures after whitening. It also suggests that modelling Long Jump, Pole Vault, and 1500 metres jointly improves the kernel density estimate compared with modelling them independently. In contrast, the remaining variables can be adequately represented by their marginal density estimators. The estimated sharpening ratio (4.9 : 4.0 : 1.1) further demonstrates that SVS estimates not only the locally dependent variable subset but also the relative contribution of each variable to the extracted local variable structure.
\begin{figure}[h]
\begin{center}
    \begin{minipage}[t]{0.38\hsize}
        \center
        \captionsetup{width=.95\linewidth}
        \includegraphics[width=\textwidth]{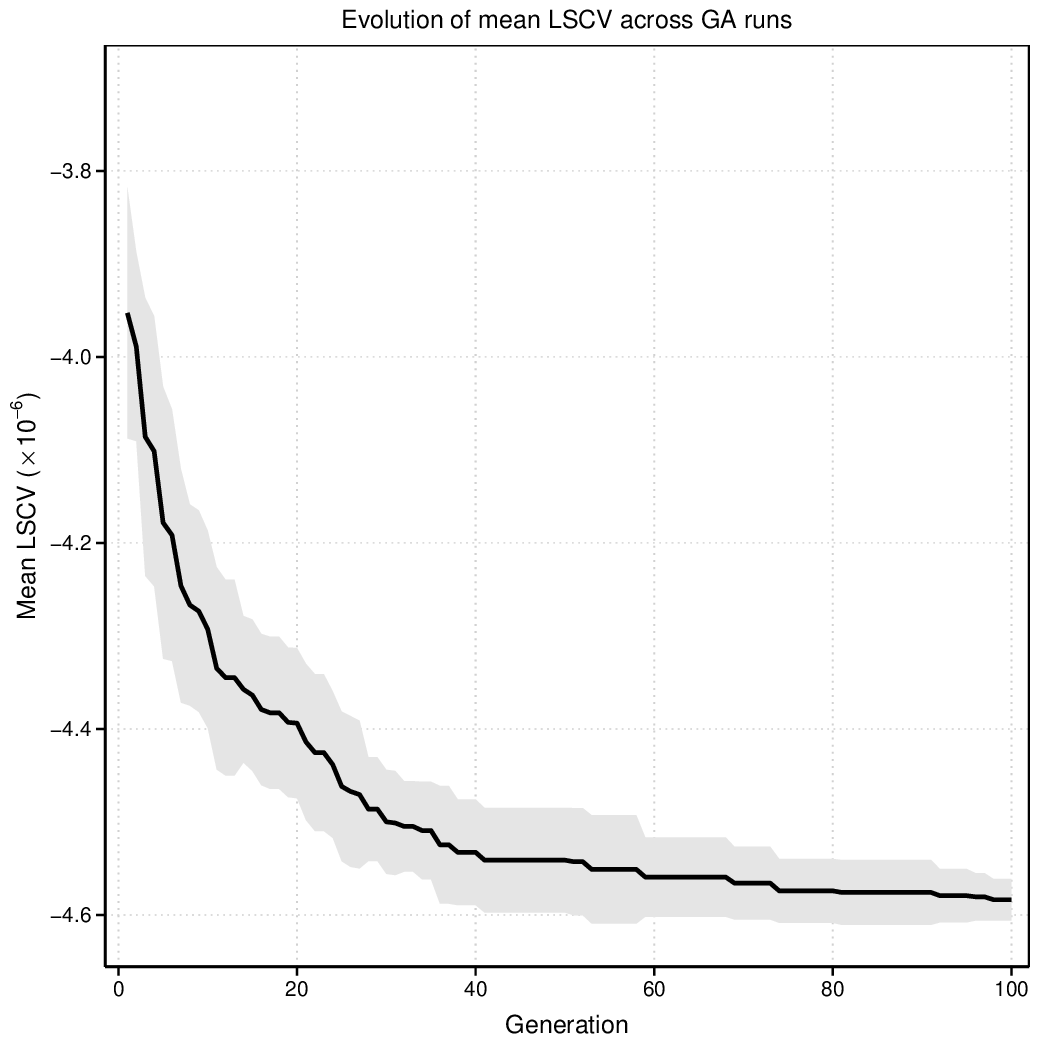}
        \mbox{(a)}
    \end{minipage}
    \begin{minipage}[t]{0.38\hsize}
        \center
        \captionsetup{width=.95\linewidth}
        \includegraphics[width=\textwidth]{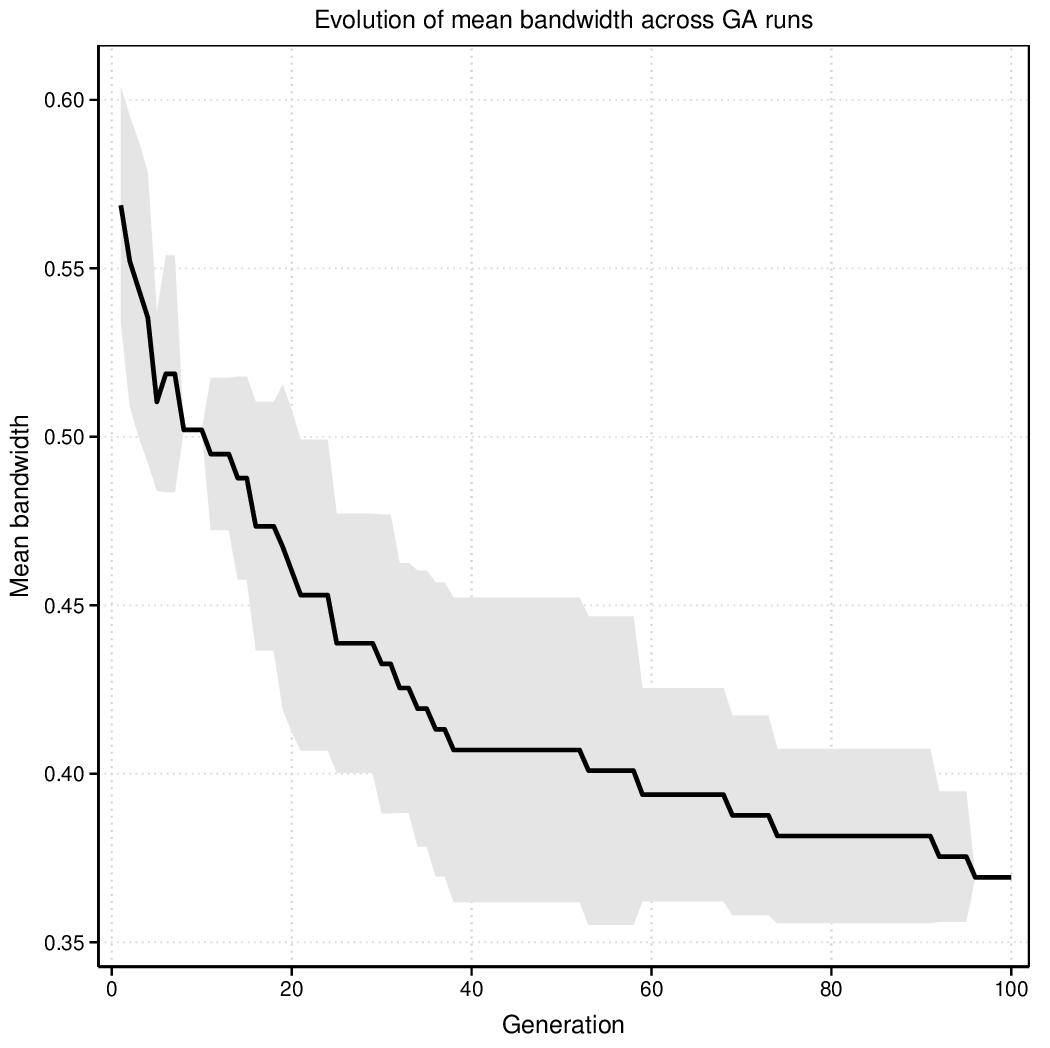}
        \mbox{(b)}
    \end{minipage}
    \begin{minipage}[t]{0.38\hsize}
        \center
        \captionsetup{width=.95\linewidth}
        \includegraphics[width=\textwidth]{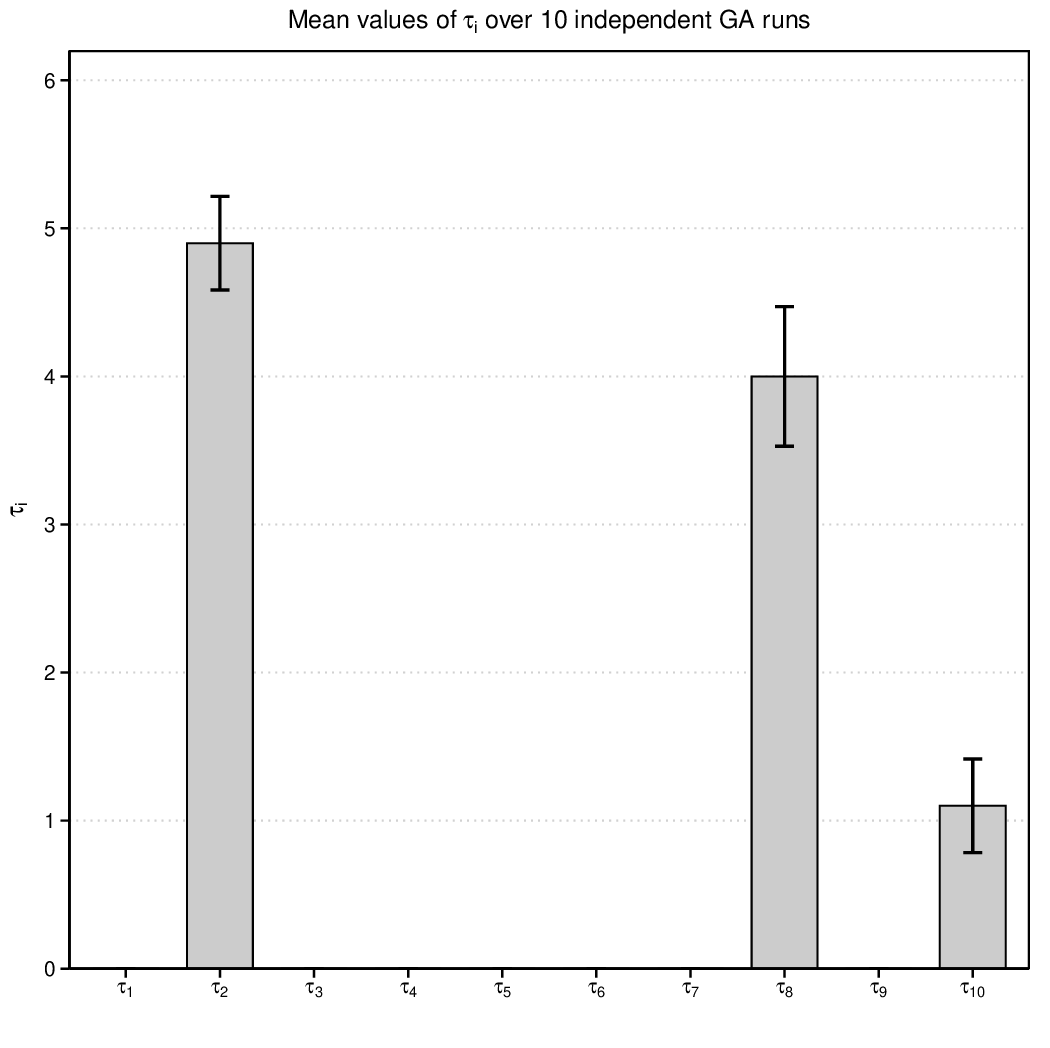}
        \mbox{(c)}
    \end{minipage}
    \begin{minipage}[t]{0.38\hsize}
        \center
        \captionsetup{width=.95\linewidth}
        \includegraphics[width=\textwidth]{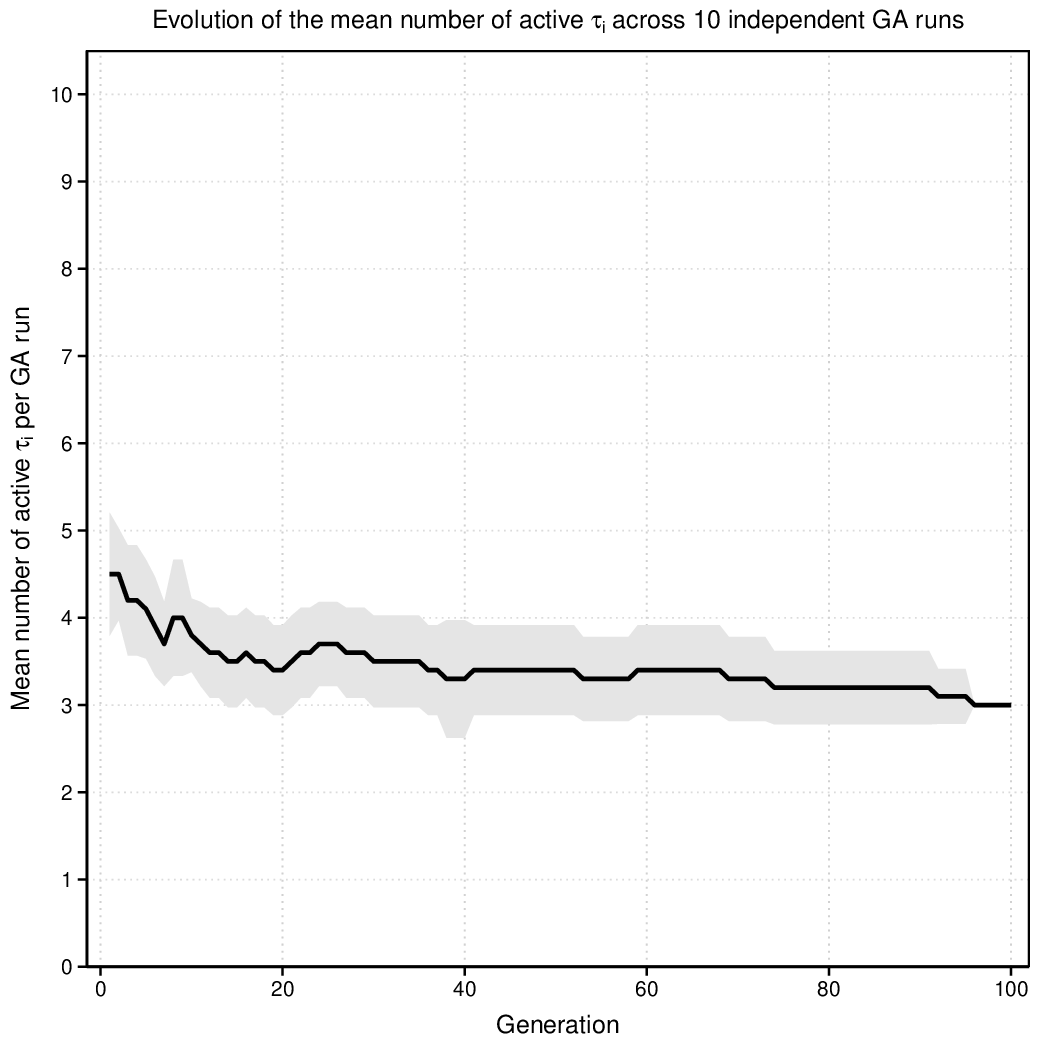}
        \mbox{(d)}
    \end{minipage}
    \begin{minipage}[t]{0.38\hsize}
        \center
        \captionsetup{width=.95\linewidth}
        \includegraphics[width=\textwidth]{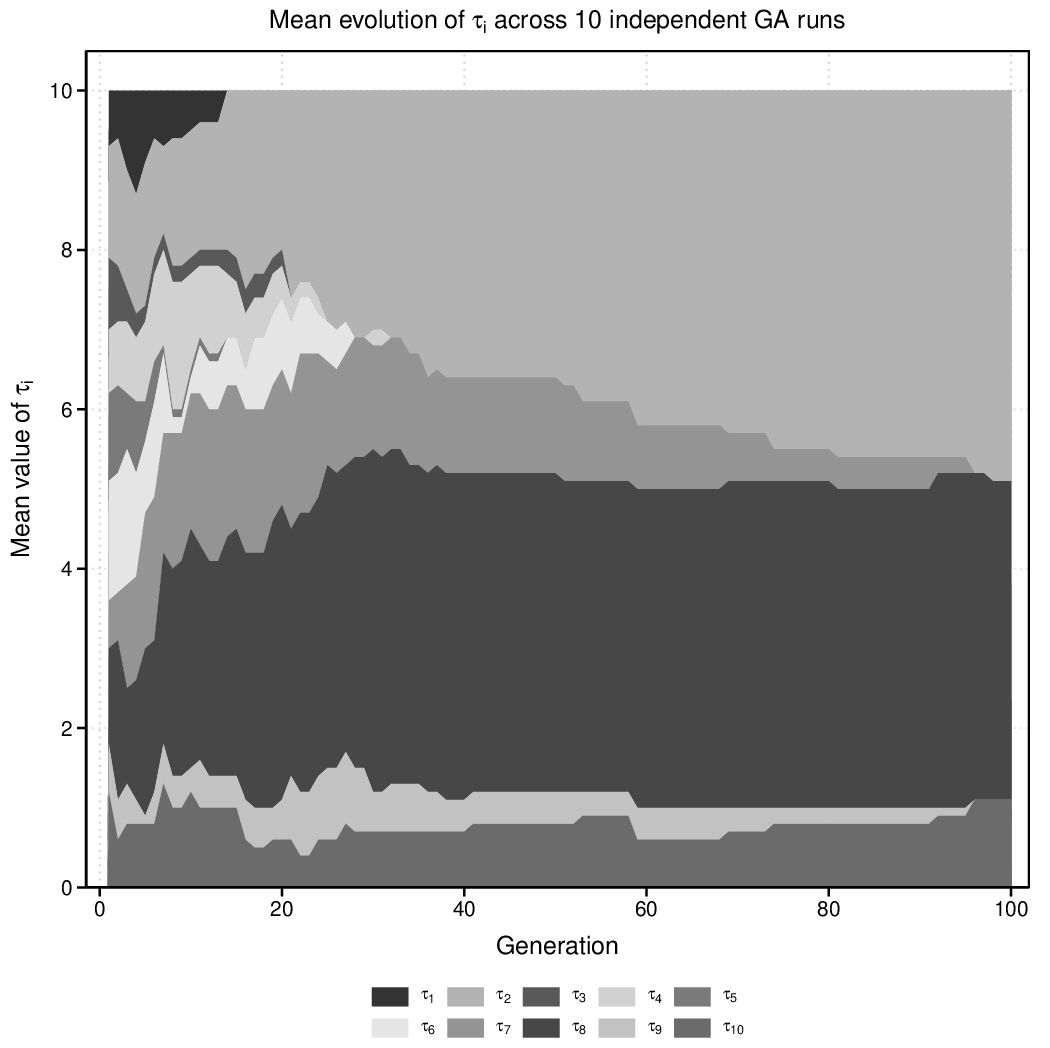}
        \mbox{(e)}
    \end{minipage}
    \begin{minipage}[t]{0.38\hsize}
        \center
        \captionsetup{width=.95\linewidth}
        \includegraphics[width=\textwidth]{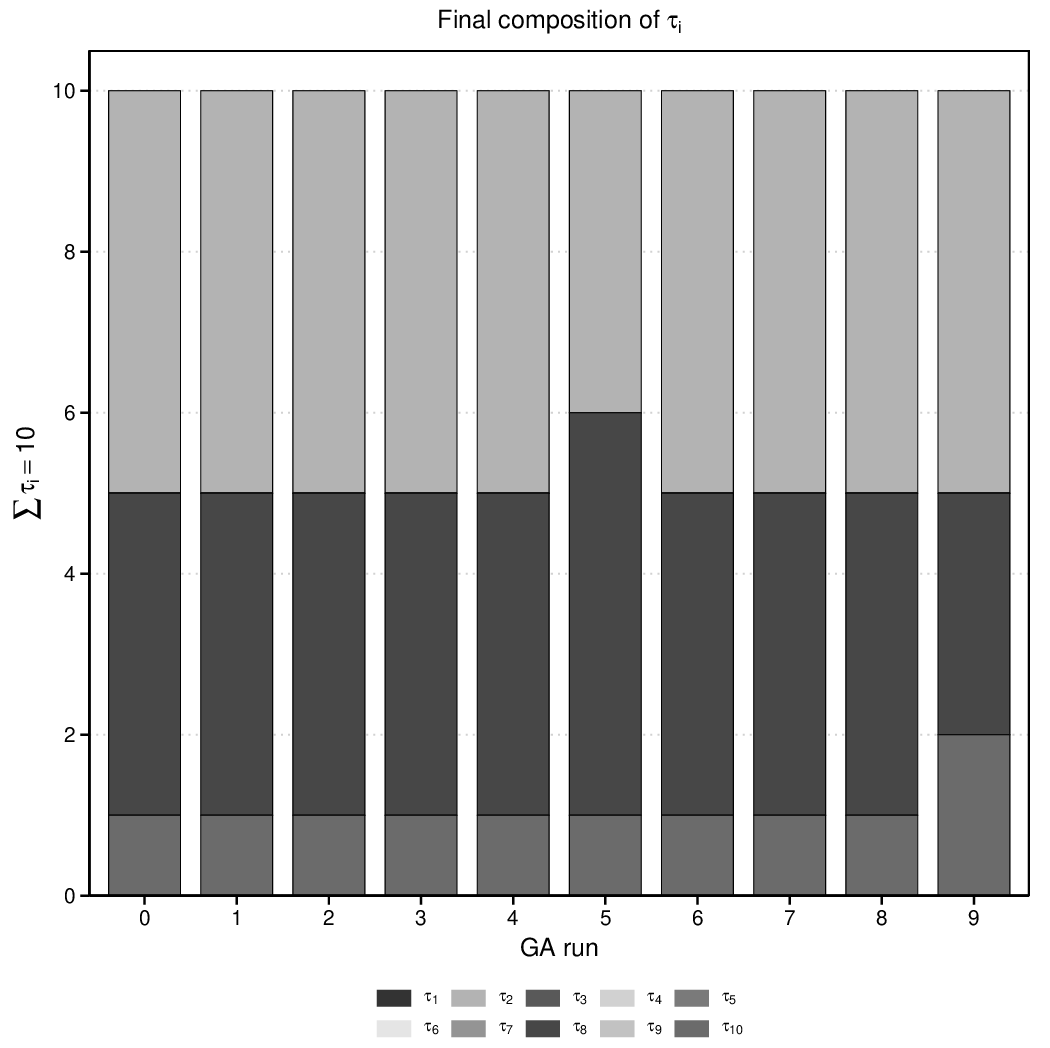}
        \mbox{(f)}
    \end{minipage}
\caption{Visualized results of SVS application to the decathlon dataset.} \label{fig.results.deca}
\end{center}
\end{figure}
\begin{table*}[tb]
\centering
\scriptsize
\setlength{\tabcolsep}{3pt}

\resizebox{\textwidth}{!}{%
\begin{tabular}{lcccccccc}
\hline
\hline
Generation
 & 1 & 10 & 20 & 30 & 40 & 50 & 75 & 100\\
\hline
$\mathrm{LSCV}^{*}$
 & -3.953 (.135)
 & -4.293 (.106)
 & -4.394 (.081)
 & -4.500 (.056)
 & -4.533 (.057)
 & -4.541 (.056)
 & -4.574 (.034)
 & -4.584 (.022)\\

Bandwidth
 & 0.569 (.035)
 & 0.502 (.000)
 & 0.460 (.048)
 & 0.433 (.044)
 & 0.407 (.045)
 & 0.407 (.045)
 & 0.382 (.026)
 & 0.369 (.000)\\

$\tau_{1}$
 & 0.700 (1.06)
 & 0.500 (1.27)
 & 0.000 (.000)
 & 0.000 (.000)
 & 0.000 (.000)
 & 0.000 (.000)
 & 0.000 (.000)
 & 0.000 (.000)\\

$\tau_{2}$
 & 1.400 (1.43)
 & 1.600 (1.26)
 & 2.000 (1.15)
 & 3.000 (1.25)
 & 3.600 (1.58)
 & 3.600 (1.58)
 & 4.500 (.850)
 & 4.900 (.316)\\

$\tau_{3}$
 & 0.900 (1.29)
 & 0.200 (.632)
 & 0.200 (.632)
 & 0.000 (.000)
 & 0.000 (.000)
 & 0.000 (.000)
 & 0.000 (.000)
 & 0.000 (.000)\\

$\tau_{4}$
 & 0.800 (.919)
 & 1.200 (1.23)
 & 0.400 (.699)
 & 0.200 (.632)
 & 0.000 (.000)
 & 0.000 (.000)
 & 0.000 (.000)
 & 0.000 (.000)\\

$\tau_{5}$
 & 1.100 (1.45)
 & 0.100 (.316)
 & 0.000 (.000)
 & 0.000 (.000)
 & 0.000 (.000)
 & 0.000 (.000)
 & 0.000 (.000)
 & 0.000 (.000)\\

$\tau_{6}$
 & 1.500 (1.84)
 & 0.200 (.632)
 & 0.900 (1.91)
 & 0.000 (.000)
 & 0.000 (.000)
 & 0.000 (.000)
 & 0.000 (.000)
 & 0.000 (.000)\\

$\tau_{7}$
 & 0.600 (1.26)
 & 1.700 (1.89)
 & 1.700 (1.83)
 & 1.300 (1.34)
 & 1.200 (1.40)
 & 1.200 (1.40)
 & 0.400 (.843)
 & 0.000 (.000)\\

$\tau_{8}$
 & 1.200 (1.14)
 & 3.000 (1.15)
 & 3.700 (1.77)
 & 4.300 (.823)
 & 4.100 (.738)
 & 4.000 (.471)
 & 4.100 (.316)
 & 4.000 (.471)\\

$\tau_{9}$
 & 0.600 (.966)
 & 0.300 (.675)
 & 0.500 (.972)
 & 0.500 (.527)
 & 0.400 (.516)
 & 0.400 (.516)
 & 0.200 (.422)
 & 0.000 (.000)\\

$\tau_{10}$
 & 1.200 (1.40)
 & 1.200 (1.75)
 & 0.600 (1.07)
 & 0.700 (1.06)
 & 0.700 (.949)
 & 0.800 (.919)
 & 0.800 (.422)
 & 1.100 (.316)\\
\hline
\hline
\end{tabular}%
}
\caption{Evolution of $\tau_{i}$ over generations for the Decathlon dataset. The numbers outside and inside parentheses denote the mean and standard deviation (S.D.), respectively. The means are reported to three decimal places, whereas the S.D.s are reported to three significant digits, with leading zeros omitted. $\mathrm{LSCV}^{*}$ values are multiplied by $10^{6}$.}
\label{tab.deca.results}
\end{table*}
\section{Discussion} \label{discussion}

This study proposes SVS to mitigate the curse of dimensionality in high-dimensional settings. Instead of estimating the full joint density directly, the proposed estimator is constructed as the product of a joint density over a selected subset of variables and marginal densities for the remaining variables. We derive the asymptotic bias, variance, AMISE, and asymptotic normality of the estimator for a fixed active index set and sharpening configuration. The theoretical results show that its performance is determined by a trade-off between approximation error and variance reduction through the choice of the active index subset. To identify suitable subsets of variables and bandwidth efficiently, we develop a GA that searches over candidate variable combinations and associated bandwidth parameters. The present theoretical analysis does not account for the additional randomness induced by the data-driven selection of $S$, $\tau$, and the bandwidth. Establishing the asymptotic properties of the fully adaptive estimator obtained after LSCV and GA selection remains an important topic for future work.

We conduct simulation studies under two scenarios: one in which some variables are mutually correlated while the others are independent, and another in which all variables are correlated. Simulation results show that the proposed estimator can accurately identify the underlying dependence structure among correlated variables and outperforms the full-joint KDE in terms of ISE. Even when all variables are mutually correlated, SVS outperforms the full-joint KDE at both sample sizes, suggesting that the variance reduction achieved through the sparse representation can outweigh the loss due to approximation even when the sparse representation does not exactly reproduce the true density. Compared with PPDE, SVS outperforms it in terms of ISE in the first scenario. However, this comparison should be regarded as being for reference only, because PPDE is primarily designed for densities that can be represented by a small number of informative projections, whereas SVS is designed for a different purpose explained as above.

The proposed SVS framework is closely related to nonparametric density estimation based on graphical models (Schwaighofer et al., \citeyear{Schwaighofer.2007}; Liu et al., \citeyear{Liu.2011}). Schwaighofer et al. (\citeyear{Schwaighofer.2007}) allow flexible decomposable structures with multiple clique and separator components; although more general in terms of factorization, their primary objective is graphical structure learning rather than mitigating the curse of dimensionality. Liu et al. (\citeyear{Liu.2011}) instead restrict the structure to a forest, reducing density estimation to univariate and bivariate components. From this perspective, SVS adopts a different restriction, consisting of one joint component of size \(|S|\leq b\) and marginal components for the remaining variables, explicitly balancing approximation and estimation errors under the curse of dimensionality. Its factorization may therefore favor dependence structures different from those suited to forest-based models.

Another important issue for SVS is how to determine the tuning parameter $b$, which can be interpreted as the total sharpening budget. In the present numerical study, we adopt a two-mode scheme with $b = d$ and $b = d/2$ as representative settings. In principle, we conjecture that the optimal value of $b$ is a monotonically increasing function of the sample size $N$, that is, $b=b(N)$, reflecting the trade-off between model complexity and estimation accuracy, because a larger sample size allows a larger sharpening budget to be accommodated without substantially increasing the estimation variance. Although the optimal function of $b=b(N)$ should ideally be determined, we adopt the present two-mode scheme for simplicity.
\clearpage
\section*{Appendix} \label{Appendix}
\subsection*{Proof of Theorem~\ref{Theorem.E.SVS}}
We denote $\mathbf{X}_{i, S}^{\top} = (X_{i,j})_{j \in S}$ to be the subvector of $\mathbf{X}_{i}^{\top}$ over the indices set $S$. Then, we obtain
\begin{eqnarray} \label{proof.SVS}
\lefteqn{E \Bigl[\widehat{q_{S}}(\mathbf{x}) \Bigr]} \nonumber \\
&=& E \Biggl[ \Biggl( \frac{1}{N} \sum_{i_{0} = 1}^{N} K_{\mathbf{H}_{S}}(\mathbf{x}_{S} - \mathbf{X}_{i_{0}, S}) \Biggr) \prod_{r=1}^{|\bar{S}|} \Biggl( \frac{1}{N} \sum_{i_{r} = 1}^{N} k_{h_{j_{r}}}(x_{j_{r}} - {X}_{i_{r}, j_{r}}) \Biggr) \Biggr] \nonumber \\
&=& \frac{1}{N^{|\bar{S}| + 1}} \sum_{i_{0} = 1}^{N} \sum_{i_{1} = 1}^{N} \cdots \sum_{i_{|\bar{S}|} = 1}^{N} E \Biggl[ K_{\mathbf{H}_{S}}(\mathbf{x}_{S} - \mathbf{X}_{i_{0}, S}) \prod_{r=1}^{|\bar{S}|} k_{h_{j_{r}}}(x_{j_{r}} - {X}_{i_{r}, j_{r}}) \Biggr]. \nonumber
\end{eqnarray}
The last equality can be decomposed into two terms $E_1$ and $E_2$, where $E_1$ corresponds to the case in which the joint and marginal components share no data points, meaning that the indices $i_{0}, i_{1}, \ldots, i_{|\bar{S}|}$ are all distinct, and $E_2$ corresponds to the case in which they share at least one data point, meaning that at least two of these indices coincide. Hence, we obtain
\begin{eqnarray} \label{proof.SVS}
\lefteqn{\frac{1}{N^{|\bar{S}| + 1}} \sum_{i_{0} = 1}^{N} \sum_{i_{1} = 1}^{N} \cdots \sum_{i_{|\bar{S}|} = 1}^{N} E \Bigl[ K_{\mathbf{H}_{S}}(\mathbf{x}_{S} - \mathbf{X}_{i_{0}, S}) \prod_{r=1}^{|\bar{S}|} k_{h_{j_{r}}}(x_{j_{r}} - {X}_{i_{r}, j_{r}}) \Bigr]} \nonumber \\
&=& \frac{1}{N^{|\bar{S}| + 1}} \Biggl[ \sum_{( i_{0}, i_{1}, ..., i_{|\bar{S}|}) \in \mathcal{C}} E \Bigl[ K_{\mathbf{H}_{S}}(\mathbf{x}_{S} - \mathbf{X}_{i_{0}, S}) \prod_{r=1}^{|\bar{S}|} k_{h_{j_{r}}}(x_{j_{r}} - {X}_{i_{r}, j_{r}}) \Bigr] \nonumber \\
& & + \sum_{( i_{0}, i_{1}, ..., i_{|\bar{S}|} ) \notin \mathcal{C}} E \Bigl[ K_{\mathbf{H}_{S}}(\mathbf{x}_{S} - \mathbf{X}_{i_{0}, S}) \prod_{r=1}^{|\bar{S}|} k_{h_{j_{r}}}(x_{j_{r}} - {X}_{i_{r}, j_{r}}) \Bigr]  \Biggr] \nonumber \\
&\equiv& \frac{1}{N^{|\bar{S}| + 1}}(E_{1} + E_{2}), \nonumber \\
\mbox{where\ }\mathcal{C} &=& \Bigl\{(i_{0}, i_{1},..., i_{|\bar{S}|}) \in \{1,2,...,N \}^{|\bar{S}| + 1}: \nexists (i_{s}, i_{t}), i_{s} = i_{t}, \forall s, t \in \{ 0, 1, ..., |\bar{S}| \} \Bigr\}. \nonumber
\end{eqnarray}
By the independence of the data, the term $E_{1}$ is written as 
\begin{eqnarray} \label{E1}
\lefteqn{E_{1} = \frac{1}{N^{|\bar{S}| + 1}} \sum_{( i_{0}, i_{1}, ..., i_{|\bar{S}|}) \in \mathcal{C}} E \Bigl[ K_{\mathbf{H}_{S}}(\mathbf{x}_{S} - \mathbf{X}_{i_{0}, S}) \prod_{r=1}^{|\bar{S}|} k_{h_{j_{r}}}(x_{j_{r}} - {X}_{i_{r}, j_{r}}) \Bigr]} \nonumber \\
&=& \frac{N(N-1)(N-2)\cdots(N - |\bar{S}|)}{N^{|\bar{S}| + 1}} E[K_{\mathbf{H}_{S}}(\mathbf{x}_{S} - \mathbf{X}_{i_{r},S})] \prod_{r=1}^{|\bar{S}|} E [ k_{h_{j_{r}}}(x_{j_{r}} - {X}_{j_{r}})] \nonumber \\
&=& \Bigl(1 + O\Bigl(\frac{1}{N}\Bigr) \Bigr) E[K_{\mathbf{H}_{S}}(\mathbf{x}_{S} - \mathbf{X}_{i_{r}, S})] \prod_{r=1}^{|\bar{S}|} E [ k_{h_{j_{r}}}(x_{j_{r}} - {X}_{j_{r}})],
\end{eqnarray}
where the terms involving $N$ is written as follows:
\begin{eqnarray}
\lefteqn{\frac{N(N-1)(N-2) \cdots (N - |\bar{S}|)}{N^{|\bar{S}| + 1}}} \nonumber \\
&=& \prod_{j=1}^{|\bar{S}|} \Bigl( 1 - \frac{j}{N} \Bigr) \nonumber \\
&=& 1 - \frac{1+2+3+\cdots+|\bar{S}|}{N}  + O\Bigl(\frac{1}{N^2}\Bigr) \nonumber \\
&=& 1 - \frac{|\bar{S}|(|\bar{S}|+1)}{2N}  + O \Bigl(\frac{1}{N^2}\Bigr) \nonumber \\
&=& 1 + O \Bigl( \frac{1}{N} \Bigr). \nonumber
\end{eqnarray}
Substituting
\begin{eqnarray} \label{E.jnt}
\lefteqn{E[K_{\mathbf{H}_{S}}(\mathbf{x}_{S} - \mathbf{X}_{i_{r},S})]} \nonumber \\
&=&  \int K_{\mathbf{H}_{S}}(\mathbf{x}_{S} - \mathbf{u}_{S})f_{S}(\mathbf{u}_{S})d\mathbf{u}_{S} \nonumber \\
&=& f_{S}(\mathbf{x}_{S}) + \frac{h_{S}^{2}}{2} \Biggl[ \sum_{j \in S}\mu_{2,j}(K_{S}){\frac{\partial^{2} f_{S}(\mathbf{x}_{S})}{\partial {x}_{j}^{2}}} \Biggl] + o(h_{S}^{2}) \nonumber
\end{eqnarray}
and
\begin{eqnarray} \label{E.mrg}
\lefteqn{E[k_{h_{k}}({x}_{k} - {X}_{j_{r}})]} \nonumber \\
&=&  \int K_{h_{k}}(x_{k} - u_{k})f_{k}(u_{k})du_{k} \nonumber \\
&=& f_{k}({x}_{k}) + \frac{h_{k}^{2}}{2} \mu_{2}(k)f_{k}^{(2)}(x_{k}) + o(h_{k}^{2}) \nonumber
\end{eqnarray}
for (\ref{E1}) (see e.g. Wand and Jones \citeyear{Wand.1995}, p.97), we obtain
\begin{eqnarray} \label{E1.last}
\lefteqn{E_{1} = q_{S}(\mathbf{x}) + \frac{h_{S}^{2}}{2} \Biggl[ \sum_{j \in S} \mu_{2,j}(K_{S}) \frac{\partial^2 f_{S}(\mathbf{x}_{S})}{\partial x_{j}^{2}} \Biggr] \prod_{k \in \bar{S}}f_{k}(x_{k})} \nonumber \\
& & \ \ \ \ + \frac{\mu_{2}(k)}{2} f_{S}(\mathbf{x}_{S}) \Biggl[ \sum_{k \in \bar{S}} h_{k}^{2}f_{k}^{(2)}(x_{k}) \prod_{l \in \bar{S}, l \neq k} f_{l}(x_{l}) \Biggr] + o(h_{S}^{2}) + o \Biggl(\sum_{k \in \bar{S}}h_{k}^{2} \Biggr) + O\Bigl(\frac{1}{N} \Bigr). \nonumber
\end{eqnarray}
In contrast, in the lack of the independence of the data, the upper bound of the term $E_{2}$ is written as,
\begin{eqnarray}
E_{2} &=& \frac{1}{N^{|\bar{S}|}} \sum_{( i_{0}, i_{1}, ..., i_{|\bar{S}|}) \notin \mathcal{C}} E \Bigl[ K_{\mathbf{H}_{S}}(\mathbf{x}_{S} - \mathbf{X}_{i_{0}, S}) \prod_{r=1}^{|\bar{S}|} k_{h_{j_{r}}}(x_{j_{r}} - {X}_{i_{r}, j_{r}}) \Bigr] \nonumber \\
&\le& |E_{2}| \nonumber \\
&\le& \frac{N^{|\bar{S}| + 1} - N(N-1)(N-2) \cdots (N - |\bar{S}|)}{N^{|\bar{S}| + 1}} \sup_{( i_{0}, i_{1}, ..., i_{|\bar{S}|}) \notin \mathcal{C}} \Biggl| E \Bigl[ K_{\mathbf{H}_{S}}(\mathbf{x}_{S} - \mathbf{X}_{i_{0}, S}) \prod_{r=1}^{|\bar{S}|} k_{h_{j_{r}}}(x_{j_{r}} - {X}_{i_{r}, j_{r}}) \Bigr] \Biggr|. \nonumber
\end{eqnarray}
Here, the term containing $N$ is written as
\begin{eqnarray}
\lefteqn{\frac{N^{|\bar{S}| + 1} - N(N-1)(N-2) \cdots (N - |\bar{S}|)}{N^{|\bar{S}| + 1}}} \nonumber \\
&=& 1 - \prod_{j=1}^{|\bar{S}|} \Bigl( 1 - \frac{j}{N} \Bigr) \nonumber \\
&=& \frac{1+2+3+\cdots+|\bar{S}|}{N}  + O\Bigl(\frac{1}{N^2}\Bigr) \nonumber \\
&=& \frac{|\bar{S}|(|\bar{S}| + 1)}{2N}  + O \Bigl(\frac{1}{N^2}\Bigr) \nonumber \\
&=& O \Bigl( \frac{1}{N} \Bigr). \nonumber
\end{eqnarray}
The term involving $\sup$ is written as
\begin{eqnarray}
\lefteqn{\sup_{( i_{0}, i_{1}, ..., i_{|\bar{S}|}) \notin \mathcal{C}} \Biggl| E \Bigl[ K_{\mathbf{H}_{S}}(\mathbf{x}_{S} - \mathbf{X}_{i_{0}, S}) \prod_{r=1}^{|\bar{S}|} k_{h_{j_{r}}}(x_{j_{r}} - {X}_{i_{r}, j_{r}}) \Bigr] \Biggr|}     \nonumber \\
& & \le \sup_{( i_{0}, i_{1}, ..., i_{|\bar{S}|}) \notin \mathcal{C}} \int \cdots \int \Bigl| K_{\mathbf{H}_{S}}(\mathbf{x}_{S} - \mathbf{u}_{i_{0}, S}) \prod_{r=1}^{|\bar{S}|} k_{h_{j_{r}}}(x_{j_{r}} - {u}_{w_{r}, j_{r}}) \Bigr| \prod_{l=1}^{q} f(\mathbf{u}_{v_{l}}) d\mathbf{u}_{v_{1}} d\mathbf{u}_{v_{2}} \cdots d\mathbf{u}_{v_{q}}, \nonumber
\end{eqnarray}
where $\{v_{1}, v_{2}, \cdots, v_{q} \} \subset \{ i_{0}, i_{1}, i_{2}, \cdots, i_{|\bar{S}|} \}$ denotes the set of distinct indices appearing in $(i_{0}, i_{1}, i_{2}, \cdots, i_{|\bar{S}|})$, and each $\mathbf{u}_{v_{l}} \in \mathbb{R}^{d}$ is a $d$-dimensional integration variable corresponding to the data $\mathbf{X}_{v_{l}}$. Here $u_{w_{r}, j_{r}}$ denotes the $j_{r}$-th component of the vector $\mathbf{u}_{w_{r}}$, where $w_{r} \in \{ v_{1}, v_{2}, ..., v_{q} \}$ is the distinct index corresponding to $i_{r}$. Since $\mathbf{X}_{1}, \mathbf{X}_{2}, ..., \mathbf{X}_{N}$ are i.i.d., the joint density of the distinct random vectors is given by $\prod_{l=1}^{q}f(u_{v_{l}})$. By {\bf{A3}}, all marginal densities of $f$ are uniformly bounded by a constant $M_{f}$. Since each distinct sample index is associated with a subset of coordinates, integration over the remaining coordinates yields the corresponding marginal density. Thus, by {\bf{A4}}, the expectation is bounded by
\begin{eqnarray}
\lefteqn{\sup_{( i_{0}, i_{1}, ..., i_{|\bar{S}|}) \notin \mathcal{C}} \int \cdots \int \Bigl| K_{\mathbf{H}_{S}}(\mathbf{x}_{S} - \mathbf{u}_{i_{0}, S}) \prod_{r=1}^{|\bar{S}|} k_{h_{j_{r}}}(x_{j_{r}} - {u}_{w_{r}, j_{r}}) \Bigr| \prod_{l=1}^{q} f(\mathbf{u}_{v_{l}}) d\mathbf{u}_{v_{1}} d\mathbf{u}_{v_{2}} \cdots d\mathbf{u}_{v_{q}}} \nonumber \\
& & \le M_{f}^{q} \| K_{S} \|_{1} \| k \|_{1}^{|\bar{S}|} = O ( 1 ), \ \ \ \ \ \ \ \ \ \ \ \ \ \ \ \ \ \ \ \ \ \ \ \ \ \ \ \ \ \ \ \ \ \ \ \ \ \ \ \ \ \ \ \ \ \ \ \ \ \ \ \ \ \ \ \ \ \ \ \ \ \ \ \ \ \ \ \ \ \ \nonumber
\end{eqnarray}
where $\int_{\mathbb{R}^{|S|}} |K_{\mathbf{S}}(\mathbf{x}_{S} - \mathbf{u}_{S})| d\mathbf{u}_{S} = \int_{\mathbb{R}^{|S|}} |K_{S}(\mathbf{t}_{S})| d\mathbf{t}_{S} < \| K_{S} \|_{1} < \infty$ and $\int_{\mathbb{R}} |k_{h_{j}}({x}_{j} - {u})| du = \int_{\mathbb{R}} |k(t)| dt < \| k \|_{1} < \infty$ by assumptions {\bf{A3-4}}. Hence, the term $E_{2}$ is written as
\begin{eqnarray}
E_{2} = O \Bigl( \frac{1}{N} \Bigr). \nonumber
\end{eqnarray}
Combining $E_{1}$ and $E_{2}$, we complete the proof.
\hfill $\Box$
\subsection*{Proof of Theorem~\ref{Theorem.Var.SVS}}
Write
\begin{eqnarray}
\widehat f_{S}(\mathbf{x}_{S}) &=& f_{S}(\mathbf{x}_{S}) + \Delta_{S}(\mathbf{x}_{S}), \nonumber
\end{eqnarray}
and, for each $j \in \bar{S}$,
\begin{eqnarray}
\widehat f_{j}(x_{j}) &=& f_{j}(x_{j}) + \Delta_{j}(x_{j}), \nonumber
\end{eqnarray}
where
\begin{eqnarray}
\Delta_{S}(\mathbf{x}_{S}) &=& \widehat f_{S}(\mathbf{x}_{S}) - f_{S}(\mathbf{x}_{S}) \ \ \mbox{and} \ \ \Delta_{j}(x_{j}) = \widehat f_{j}(x_{j}) - f_{j}(x_{j}). \nonumber
\end{eqnarray}
Then, we obtain
\begin{eqnarray}
\widehat{q_{S}}(\mathbf{x}) &=& \bigl(f_{S}(\mathbf{x}_{S}) + \Delta_{S}(\mathbf{x}_{S}) \bigr) \prod_{k \in \bar{S}} \bigl(f_{k}(x_{k}) + \Delta_{k}(x_{k}) \bigr) \nonumber \\
&=& q_{S}(\mathbf{x}) + \Delta_{S}(\mathbf{x}_{S}) \prod_{k \in \bar{S}} f_{k}(x_{k}) + \Biggl( \prod_{k \in \bar{S}} f_{k}(x_{k}) \Biggr) \sum_{j \in \bar{S}} \Delta_{j}(x_{j}) \frac{f_{S}(\mathbf{x}_{S})}{f_{j}({x}_{j})} + \mbox{Rem}(\mathbf{x}), \phantomsection \label{g_SVS_expansion} \\
\mbox{Rem}(\mathbf{x}) &=& q_{S}(\mathbf{x}) \left[ \left(1 + \frac{\Delta_{S}(\mathbf{x}_{S})}{f_{S}(\mathbf{x}_{S})} \right) \prod_{j \in \bar{S}} \left(1 + \frac{\Delta_{j}(x_{j})}{f_{j}(x_{j})}\right) \right.  \left. -1 - \frac{\Delta_{S}(\mathbf{x}_{S})}{f_{S}(\mathbf{x}_{S})} - \sum_{j \in \bar{S}} \frac{\Delta_{j}(x_{j})}{f_{j}(x_{j})} \right], \nonumber
\end{eqnarray}
where $\mbox{Rem}(\mathbf{x})$ collects all terms involving at least two $\Delta$'s. Subtracting the expectation of each side of \eqref{g_SVS_expansion} from the corresponding side of \eqref{g_SVS_expansion}, we obtain
\begin{eqnarray} \label{q.centered}
\widehat{q_{S}}(\mathbf{x}) - E[ \widehat{q_{S}}(\mathbf{x})] &=& A_{S}(\mathbf{x}) + \sum_{j \in \bar{S}} A_{j}(\mathbf{x}) + \mbox{Rem}^{*}(\mathbf{x}),
\end{eqnarray}
where
\begin{eqnarray}
A_{S}(\mathbf{x}) &=& \Biggl( \prod_{k \in \bar{S}} f_{k}(x_{k}) \Biggr) \bigl(\widehat f_{S}(\mathbf{x}_{S}) - E[\widehat f_{S}(\mathbf{x}_{S})] \bigr), \nonumber \\
A_{j}(\mathbf{x}) &=& \Biggl( \prod_{k \in \bar{S}} f_{k}(x_{k}) \Biggr) \bigl(\widehat f_{j}(x_{j}) - E[\widehat f_{j}(x_{j})] \bigr) \frac{f_{S}(\mathbf{x}_{S})}{f_{j}(x_{j})}, \qquad j \in \bar{S}, \nonumber \\
\mbox{Rem}^{*}(\mathbf{x}) &=& \mbox{Rem}(\mathbf{x}) - E[\mbox{Rem}(\mathbf{x})]. \nonumber 
\end{eqnarray}
Hence,
\begin{eqnarray}
\mbox{Var} \!\left[ \widehat{q_{S}}(\mathbf{x}) \right] &=& \mbox{Var} \Biggl[ A_{S}(\mathbf{x}) + \sum_{j \in \bar{S}} A_{j}(\mathbf{x}) + \mbox{Rem}^{*}(\mathbf{x}) \Biggr]. \nonumber
\end{eqnarray}
Since $\widehat f_{S}(\mathbf{x}_{S})$ is a $|S|$-dimensional KDE with bandwidth $h_{S}$, its variance is given by (see e.g., Wand and Jones, 1995, p.97)
\begin{eqnarray}
\mathrm{Var}\!\left[ \widehat{f_{S}}(\mathbf{x}_{S}) \right] &=& \frac{R(K_{S})\,f_{S}(\mathbf{x}_{S})}{N h_{S}^{|S|}} + o\! \left(\frac{1}{N h_{S}^{|S|}} \right).  \nonumber
\end{eqnarray}
Therefore,
\begin{eqnarray} \label{eq:var_A0_barS}
\mbox{Var}[A_{S}(\mathbf{x})] &=& \Biggl( \prod_{k \in \bar{S}} f_{k}(x_{k})^{2} \Biggr) \mbox{Var} \!\left[ \widehat f_{S}(\mathbf{x}_{S}) \right] \nonumber \\
&=& \Biggl( \prod_{k \in \bar{S}} f_{k}(x_{k})^{2} \Biggr) \frac{R(K_{S}) f_{S}(\mathbf{x}_{S})}{N h_{S}^{|S|}} + o \!\left( \frac{1}{N h_{S}^{|S|}} \right).
\end{eqnarray}
Similarly, for each $j \in \bar{S}$,
\begin{eqnarray}
\mbox{Var} \!\left[ \widehat f_{j}(x_{j}) \right] &=& \frac{R(k) f_{j}(x_{j})}{N h_{j}} + o \!\left( \frac{1}{N h_{j}} \right),  \nonumber
\end{eqnarray}
and hence
\begin{eqnarray} \label{eq:var_Aj_barS}
\mbox{Var}[A_{j}(\mathbf{x})] &=& \frac{f_{S}(\mathbf{x}_{S})^2}{f_{j}(x_{j})^2} \Biggl( \prod_{k \in \bar{S}} f_{k}(x_{k})^2 \Biggr) \mbox{Var} \!\left[ \widehat f_{j}(x_{j}) \right] \nonumber \\
&=& \frac{f_{S}(\mathbf{x}_{S})^2}{f_{j}(x_{j})} \Biggl( \prod_{k \in \bar{S}} f_{k}(x_{k})^2 \Biggr)\frac{R(k)}{N h_{j}} + o \! \left( \frac{1}{N h_{j}} \right).
\end{eqnarray}
For $j \in \bar{S}$ and any sample index numbers $a_{0}$, $a_{1} \in \{1, 2, \cdots N \}, a_{0} \neq a_{1}$, we obtain
\begin{eqnarray} \label{cov_A0_Aj}
\lefteqn{\mbox{Cov}(A_{S}(\mathbf{x}), A_{j}(\mathbf{x}))} \nonumber \\
&=& \Biggl[ \frac{f_{S}(\mathbf{x}_{S})}{f_{j}(x_{j})} \prod_{k \in \bar{S}} f_{k}(x_{k})^2 \Biggr] \mbox{Cov}\Bigl(\widehat{f_{S}}(\mathbf{x}_{S}),\widehat{f_{j}}(x_{j}) \Bigr) \nonumber \\
&=& \mbox{Const.} \times \mbox{Cov}\Bigl(\frac{1}{N} \sum_{i_{0}=1}^{N} K_{\mathbf{H}_{S}}(\mathbf{x}_{S} - \mathbf{X}_{i_{0}, S}) , \frac{1}{N} \sum_{i_{1}=1}^{N} k_{{h}_{j}}(x_{j} - {X}_{i_{1}, j}) \Bigr)  \nonumber \\
&\propto& \frac{1}{N^{2}} \sum_{i_{0}=1}^{N} \sum_{i_{1}=1}^{N} \mbox{Cov} \Bigl( K_{\mathbf{H}_{S}}(\mathbf{x}_{S} - \mathbf{X}_{i_{0}, S}), k_{{h}_{j}}(x_{j} - {X}_{i_{1}, j}) \Bigr) \nonumber \\
&=& \frac{1}{N^{2}} \Biggl( N \mbox{Cov} \Bigl( K_{\mathbf{H}_{S}}(\mathbf{x}_{S} - \mathbf{X}_{a_{0}, S}), k_{{h}_{j}}(x_{j} - {X}_{a_{0}, j}) \Bigr) \nonumber \\
&& \ \ \ \ \ \ \ \ \ \ \ \ \ \ \  + N(N-1) \mbox{Cov} \Bigl( K_{\mathbf{H}_{S}}(\mathbf{x}_{S} - \mathbf{X}_{a_{0}, S}), k_{{h}_{j}}(x_{j} - {X}_{a_{1}, j}) \Bigr) \Biggr) \nonumber \\
&=& \frac{1}{N} \mbox{Cov} \Bigl( K_{\mathbf{H}_{S}}(\mathbf{x}_{S} - \mathbf{X}_{a_{0}, S}), k_{{h}_{j}}(x_{j} - {X}_{a_{0}, j}) \Bigr) \nonumber \\
&\le& \frac{1}{N} \Bigl| \mbox{Cov} \Bigl( K_{\mathbf{H}_{S}}(\mathbf{x}_{S} - \mathbf{X}_{a_{0}, S}), k_{{h}_{j}}(x_{j} - {X}_{a_{0}, j}) \Bigr) \Bigr| \nonumber \\
&\le& \frac{1}{N} \Bigl( E \Bigl|K_{\mathbf{H}_{S}}(\mathbf{x}_{S} - \mathbf{X}_{a_{0}, S}) k_{{h}_{j}}(x_{j} - {X}_{a_{0}, j})\Bigr| + E \Bigl|K_{\mathbf{H}_{S}}(\mathbf{x}_{S} - \mathbf{X}_{a_{0}, S})\Bigr| E \Bigl|k_{{h}_{j}}(x_{j} - {X}_{a_{0}, j}) \Bigr| \Bigr) \nonumber \\
&\le& \frac{M_{f} + M_{f}^{2}}{N} \| K_{S} \|_{1} \| k \|_{1} \nonumber \\
&=& O \Bigl( \frac{1}{N} \Bigr), \nonumber
\end{eqnarray}
where the second from the last inequality comes from {\bf{A3-4}}.

Similarly, for any pair of indices $j_{1}, j_{2} \in \bar{S}$ such that $j_{1} \neq j_{2}$, we obtain
\begin{eqnarray} \label{cov_Aj_Ak}
\lefteqn{\mbox{Cov}(A_{j_{1}}(\mathbf{x}), A_{j_{2}}(\mathbf{x}))} \nonumber \\
&=& \Biggl[ f_{S}(\mathbf{x}_{S})^{2} \prod_{k \in \bar{S}} \frac{f_{k}(x_{k})^{2}}{f_{j_{1}}(x_{j_{1}}) f_{j_{2}}(x_{j_{2}})} \Biggr] \mbox{Cov}\Bigl(\widehat{f_{j_{1}}}(x_{j_{1}}), \widehat{f_{j_{2}}}(x_{j_{2}}) \Bigr) \nonumber \\
&=& \mbox{Const.} \times \mbox{Cov}\Bigl(\frac{1}{N} \sum_{i_{0}=1}^{N} k_{{h}_{j_{1}}}(x_{j_{1}} - {X}_{i_{0}, j_{1}}), \frac{1}{N} \sum_{i_{1}=1}^{N} k_{{h}_{j_{2}}}(x_{j_{2}} - {X}_{i_{1}, j_{2}}) \Bigr)  \nonumber \\
&\propto& \frac{1}{N^{2}} \sum_{i_{0}=1}^{N} \sum_{i_{1}=1}^{N} \mbox{Cov} \Bigl( k_{{h}_{j_{1}}}(x_{j_{1}} - {X}_{i_{0}, j_{1}}), k_{{h}_{j_{2}}}(x_{j_{2}} - {X}_{i_{1}, j_{2}}) \Bigr) \nonumber \\
&=& \frac{1}{N^{2}} \Biggl( N \mbox{Cov} \Bigl( k_{{h}_{j_{1}}}(x_{j_{1}} - {X}_{a_{0}, j_{1}}),  k_{{h}_{j_{2}}}(x_{j_{2}} - {X}_{a_{0}, j_{2}}) \Bigr) \nonumber \\
&& \ \ \ \ \ \ \ \ \ \ \ \ \ \ \ + N(N-1) \mbox{Cov} \Bigl( k_{{h}_{j_{1}}}(x_{j_{1}} - {X}_{a_{0}, j_{1}}), k_{{h}_{j_{2}}}(x_{j_{2}} - {X}_{a_{1}, j_{2}}) \Bigr) \Biggr) \nonumber \\
&=& \frac{1}{N} \mbox{Cov} \Bigl( k_{{h}_{j_{1}}}(x_{j_{1}} - {X}_{a_{0}, j_{1}}),  k_{{h}_{j_{2}}}(x_{j_{2}} - {X}_{a_{0}, j_{2}}) \Bigr) \nonumber \\
&\le& \frac{1}{N} \Bigl| \mbox{Cov} \Bigl( k_{{h}_{j_{1}}}(x_{j_{1}} - {X}_{a_{0}, j_{1}}),  k_{{h}_{j_{2}}}(x_{j_{2}} - {X}_{a_{0}, j_{2}}) \Bigr) \Bigr| \nonumber \\
&\le& \frac{1}{N} \Bigl( E \Bigl| k_{{h}_{j_{1}}}(x_{j_{1}} - {X}_{a_{0}, j_{1}})  k_{{h}_{j_{2}}}(x_{j_{2}} - {X}_{a_{0}, j_{2}}) \Bigr| + E \Bigl| k_{{h}_{j_{1}}}(x_{j_{1}} - {X}_{a_{0}, j_{1}}) \Bigr| E \Bigl| k_{{h}_{j_{2}}}(x_{j_{2}} - {X}_{a_{0}, j_{2}}) \Bigr| \Bigr) \nonumber \\
&\le& \frac{M_{f} + M_{f}^{2}}{N}  \| k \|_{1}^{2} \nonumber \\
&=& O \Bigl( \frac{1}{N} \Bigr), \nonumber
\end{eqnarray}
using {\bf{A3-4}}. Since $h_{S} \to 0$ and $h_{j} \to 0$ as $N \to \infty$, we obtain
\begin{eqnarray}
\frac{1/N}{1/(Nh^{|S|})} = h_{S}^{|S|} \to 0, \ \ \ 
\frac{1/N}{1/(Nh_{j})} = h_{j} \to 0, \ \ \  \nonumber
\end{eqnarray}
so all covariance terms are of smaller order than the leading variance terms.

By standard moment bounds for KDE and H\"older's inequality, the higher-order remainder terms, including those involving the bias components, satisfy
\begin{eqnarray} \phantomsection \label{V.Rem}
\mbox{Var} \left( \mbox{Rem}^{*}(\mathbf{x}) \right) = o \left( W_{V} \right).
\end{eqnarray}
Furthermore, by the Cauchy--Schwarz inequality,
\begin{eqnarray} \label{Cov.AS.Rem}
\lefteqn{\Bigl| \mbox{Cov} \Bigl(A_{S}(\mathbf{x}) + \sum_{j \in \bar S} A_{j}(\mathbf{x}), \mbox{Rem}^{*}(\mathbf{x}) \Bigr) \Bigr|} \nonumber \\
&\le& \mbox{Var} \Bigl( A_{S}(\mathbf{x}) + \sum_{j \in \bar S} A_{j}(\mathbf{x}) \Bigr)^{\frac{1}{2}} \mbox{Var} \left( \mbox{Rem}^{*}(\mathbf{x}) \right)^{\frac{1}{2}} = \bigl(O(W_{V})o(W_{V}) \bigr)^{\frac{1}{2}} = o(W_{V}).
\end{eqnarray}
Combining \eqref{eq:var_A0_barS}, \eqref{eq:var_Aj_barS}, \eqref{V.Rem} and \eqref{Cov.AS.Rem} completes the proof.
\hfill $\Box$
\subsection*{Proof of Theorem~\ref{Theorem.MISE.bound}}
The MISE can be decomposed as follows:
\begin{eqnarray}
\lefteqn{MISE \bigl(\widehat{q_{S}}(\mathbf{x}), f(\mathbf{x}) \bigr)} \nonumber \\
&=& \int_{\mathbb{R}^{d}} E \Bigl[ \bigl( \widehat{q}_{S}(\mathbf{x}) - E[\widehat{q}_{S}(\mathbf{x})] + E[\widehat{q}_{S}(\mathbf{x})] - q_{S}(\mathbf{x}) + q_{S}(\mathbf{x}) - f(\mathbf{x}) \bigl)^{2} \Bigl] d\mathbf{x} \nonumber\\
&=& \int_{\mathbb{R}^{d}} \mbox{Var}[\widehat{q}_{S}(\mathbf{x})] d\mathbf{x} + \int_{\mathbb{R}^{d}} \bigl( E[\widehat{q}_{S}(\mathbf{x})] - q_{S}(\mathbf{x}) \bigr)^{2} d\mathbf{x}  \nonumber \\
& & + 2 \int_{\mathbb{R}^{d}} \bigl( E[\widehat{q}_{S}(\mathbf{x})] - q_{S}(\mathbf{x}) \bigr) \bigl( q_{S}(\mathbf{x}) - f(\mathbf{x}) \bigr) d\mathbf{x} + \int_{\mathbb{R}^{d}} \bigl( q_{S}(\mathbf{x}) - f(\mathbf{x}) \bigr)^{2} d\mathbf{x}. \nonumber
\end{eqnarray}
Motivated by this decomposition, the AMISE is defined by replacing the bias and variance components with their leading terms:
\begin{eqnarray} \label{AMISE_q}
\lefteqn{AMISE \bigl(\widehat{q_{S}}(\mathbf{x}), f(\mathbf{x}) \bigr)} \nonumber \\
&=& \int_{\mathbb{R}^{d}} LV(\mathbf{x})d\mathbf{x} + \int_{\mathbb{R}^{d}} LB(\mathbf{x})^{2} d\mathbf{x} \nonumber \\
& & + 2 \int_{\mathbb{R}^{d}} LB(\mathbf{x}) \bigl( q_{S}(\mathbf{x}) - f(\mathbf{x}) \bigr) d\mathbf{x} + \int_{\mathbb{R}^{d}} \bigl( q_{S}(\mathbf{x}) - f(\mathbf{x}) \bigr)^{2} d\mathbf{x}.
\end{eqnarray}
From Theorem~\ref{Theorem.E.SVS}, the leading bias term satisfies $O \Bigl( h_{S}^{2} + \sum_{j \in \bar{S}} h_{j}^{2} \Bigr)$.
Expanding the square yields
\begin{eqnarray}
\Biggl(h_{S}^{2} + \sum_{j \in \bar{S}} h_{j}^{2} \Biggr)^{2} &=& h_{S}^{4} + 2 h_{S}^{2} \sum_{j \in \bar{S}} h_{j}^{2} + \Biggl( \sum_{j \in \bar{S}} h_{j}^{2} \Biggr)^{2} \nonumber \\
&\le& h_{S}^{4} + \Biggl(h_{S}^{4} + \Biggl(\sum_{j \in \bar{S}} h_{j}^{2} \Biggr)^{2} \Biggr) + \Biggl(\sum_{j \in \bar{S}} h_{j}^{2} \Biggr)^{2} \nonumber \\
&=& 2 h_{S}^{4} + 2 \sum_{j, k \in \bar{S}} h_{j}^{2} h_{k}^{2} \nonumber \\
&\le& 2h_{S}^{4} + 2 |\bar{S}| \sum_{j \in \bar{S}} h_{j}^{4} \nonumber \\
&=& O(W_{B}). \nonumber
\end{eqnarray}
By Theorems~\ref{Theorem.E.SVS}-\ref{Theorem.Var.SVS} together with {\bf{A3-4}}, the leading terms of the integrated squared bias and variance satisfy $\int_{\mathbb{R}^d} LB(\mathbf{x})^2 d\mathbf{x} = O(W_{B})$ and $\int_{\mathbb{R}^d}LV(x) d\mathbf{x} = O(W_{V})$.
Consequently,
\begin{eqnarray}
\int_{\mathbb{R}^d} LB(\mathbf{x})^2 d\mathbf{x} + \int_{\mathbb{R}^d} LV(\mathbf{x}) d\mathbf{x} = O(W_{N}). \nonumber 
\end{eqnarray}
For the cross term in (\ref{AMISE_q}), the Cauchy--Schwarz inequality followed by Young's inequality gives, for any fixed $0 < \varepsilon < 1$,
\begin{eqnarray}
\lefteqn{2 \left| \int_{\mathbb R^d} LB(\mathbf{x}) \bigl( q_{S}(\mathbf{x}) - f(\mathbf{x}) \bigr) d\mathbf{x} \right|} \nonumber \\
&\leq& 2 \left[ \int_{\mathbb R^{d}} LB(\mathbf{x})^{2} d\mathbf{x} \right]^{\frac{1}{2}} \left[ \int_{\mathbb{R}^d} \bigl( q_{S}(\mathbf{x}) - f(\mathbf{x}) \bigr)^2 d\mathbf{x} \right]^{\frac{1}{2}} \nonumber \\
&=& 2 \left[ \int_{\mathbb R^{d}} LB(\mathbf{x})^{2} d\mathbf{x} \right]^{\frac{1}{2}} \mathcal{E}(f,q_{S})^{\frac{1}{2}} \nonumber \\
&\leq& \frac{1}{\varepsilon} \int_{\mathbb R^{d}} LB(\mathbf{x})^{2} d\mathbf{x} + \varepsilon \mathcal{E}(f,q_{S}) \nonumber \\
&=& O (W_{B}) + \varepsilon \mathcal{E}(f,q_{S}). \nonumber
\end{eqnarray}
Thus, the upper bound of (\ref{AMISE_q}) is written as
\begin{eqnarray}
\lefteqn{AMISE \Bigl( \widehat q_{S}(\mathbf{x}), f(\mathbf{x}) \Bigr)} \nonumber \\
&\leq& O(W_{N}) + (1 + \varepsilon) \mathcal E(f,q_{S}) + O (W_{B}) \nonumber \\
&=& (1 + \varepsilon) \mathcal E(f,q_{S}) + O(W_{N}). \nonumber
\end{eqnarray}
The lower bound is written as
\begin{eqnarray}
\lefteqn{AMISE \Bigl( \widehat q_{S}(\mathbf{x}), f(\mathbf{x}) \Bigr)} \nonumber \\
&\ge& \mathcal E(f,q_{S}) - 2 \left| \int_{\mathbb R^d} LB(\mathbf{x}) \Bigl( q_{S}(\mathbf{x}) - f(\mathbf{x}) \Bigr) d\mathbf{x} \right| \nonumber \\
&\ge& \mathcal E(f,q_{S}) - \frac{1}{\varepsilon} \int_{\mathbb R^{d}} LB(\mathbf{x})^{2} d\mathbf{x} - \varepsilon \mathcal{E}(f,q_{S}) \nonumber \\
&=& (1 - \varepsilon) \mathcal E(f,q_{S}) - O ( W_{B}). \nonumber
\end{eqnarray}
Finally, balancing the squared-bias and variance terms in $O(W_{N})$ separately for the joint and marginal parts yields
\begin{eqnarray}
W_{N} = O \left(N^{-\frac{4}{|S| + 4}} + |\bar S|N^{-\frac{4}{5}} \right). \nonumber
\end{eqnarray}
This completes the proof.
\hfill $\Box$
\subsection*{Proof of Theorem~\ref{Theorem.AN.SVS}}
We consider the centered decomposition in (\ref{q.centered}).  For each $j \in \bar S$, Theorem~\ref{Theorem.Var.SVS} gives
\begin{eqnarray}
\mbox{Var}[A_{j}(\mathbf{x})] = O \left( \frac{1}{Nh_{j}} \right). \nonumber
\end{eqnarray}
Since $E[A_{j}(\mathbf{x})] = 0$, by $\bf{A5}$ and Chebyshev's inequality,
\begin{eqnarray}
\sqrt{Nh_{S}^{|S|}}A_{j}(\mathbf{x}) = O_{p} \left(\sqrt{\frac{h_{S}^{|S|}}{h_{j}}} \right) = o_{p}(1). \nonumber
\end{eqnarray}
Hence, the second term in the right hand side of (\ref{q.centered}) vanishes.

The standard bias and variance bounds gives
\begin{eqnarray}
E[ \Delta_{S}] = O \left( h_{S}^{2} \right),\ \ \ E[\Delta_{j}] = O \left( h_{j}^{2} \right),\ \ \ \mbox{Var}(\Delta_{S}) =O \left( \frac{1}{Nh_S^{|S|}} \right),\ \ \ \mbox{Var}(\Delta_{j}) = O \left(\frac{1}{Nh_{j}} \right). \nonumber
\end{eqnarray}
Writing each error as its expectation plus its centered component, we obtain
\begin{eqnarray}
\lefteqn{\Delta_{S} \Delta_{j}} \nonumber \\
&=& \Bigl[ E[\Delta_{S}] + (\Delta_{S} - E[\Delta_{S}]) \Bigr] \Bigl[ E[\Delta_{j}] + (\Delta_{j} - E[\Delta_{j}]) \Bigr] \nonumber \\
&=& E[\Delta_{S}] E[\Delta_{j}] + E[\Delta_{S}] ( \Delta_{j} - E[\Delta_{j}] ) + E[\Delta_{j}] ( \Delta_{S} - E[\Delta_{S}]) + (\Delta_{S} - E[\Delta_{S}])(\Delta_{j} - E[\Delta_{j}]). \nonumber
\end{eqnarray}
Taking expectations gives
\begin{eqnarray}
E[\Delta_{S} \Delta_{j}] = E[\Delta_{S}] E[\Delta_{j}] + \mbox{Cov}(\Delta_{S},\Delta_{j}). \nonumber
\end{eqnarray}
Subtracting this identity yields
\begin{eqnarray}
\lefteqn{\Delta_{S} \Delta_{j} - E[\Delta_{S} \Delta_{j}]} \nonumber\\
&=& E[\Delta_{S}] ( \Delta_{j} - E[\Delta_{j}] ) + E[\Delta_{j}] (\Delta_{S} - E[\Delta_{S}]) + (\Delta_{S} - E[\Delta_{S}]) (\Delta_{j} - E[\Delta_{j}]) - \mbox{Cov}(\Delta_{S},\Delta_{j}). \nonumber
\end{eqnarray}
By the Cauchy--Schwarz inequality,
\begin{eqnarray}
|\operatorname{Cov}(\Delta_S,\Delta_j)| \leq \sqrt{\mbox{Var}(\Delta_{S}) \mbox{Var}(\Delta_{j})} = O \left(\frac{1}{N \sqrt{h_{S}^{|S|}h_{j}}} \right). \nonumber
\end{eqnarray}
Consequently, by {\bf{A5}}, we obtain
\begin{eqnarray}
\sqrt{Nh_S^{|S|}} \Bigl( \Delta_{S} \Delta_{j} - E[\Delta_{S} \Delta_{j}] \Bigr) = O_{p} \left(h_{S}^{2} \sqrt{\frac{h_{S}^{|S|}}{h_{j}}} + h_{j}^{2} + \frac{1}{\sqrt{Nh_{j}}} \right)= o_{p}(1). \nonumber
\end{eqnarray}
By Holder's inequality and standard KDE moment bounds, all remaining higher-order centered products are negligible under {\bf{A1-5}}, yielding
\begin{eqnarray}
\sqrt{Nh_{S}^{|S|}}\operatorname{Rem}^{*}(\mathbf{x}) = o_{p}(1). \nonumber
\end{eqnarray}

By the standard asymptotic normality of the multivariate KDE,
\begin{eqnarray}
\sqrt{Nh_{S}^{|S|}} \left( \widehat f_{S}(\mathbf{x}_{S}) - E[\widehat f_{S}(\mathbf{x}_{S})] \right) \xrightarrow{d} N\left(0, f_{S}(\mathbf{x}_{S})R(K_{S}) \right). \nonumber
\end{eqnarray}
Hence, by Slutsky's theorem,
\begin{eqnarray}
\sqrt{Nh_{S}^{|S|}} \Bigl( \widehat q_{S}(\mathbf{x}) - E[\widehat q_{S}(\mathbf{x})] \Bigr) \xrightarrow{d}
N \left(0, \left(\prod_{k \in \bar{S}}f_{k}(x_{k}) \right)^2 f_{S}(\mathbf{x}_{S})R(K_{S}) \right). \nonumber
\end{eqnarray}
Finally, since
\begin{eqnarray}
\widehat{q_{S}}(\mathbf{x}) - E[\widehat{q_{S}}(\mathbf{x})] = \widehat{q_{S}}(\mathbf{x}) - q_{S}(\mathbf{x}) - \mbox{Bias}(\widehat q_{S}(\mathbf{x})), \nonumber
\end{eqnarray}
the desired result follows.

The case $|S|=1$ is excluded because $q_{S}(\mathbf{x})$ reduces to the all-marginal approximation. This completes the proof.
\hfill $\Box$

\section*{Acknowledgements}
The author gratefully acknowledges the financial support from JSPS KAKENHI 23K28043 and 26K04853.


\begin{thebibliography}{3}
\setlength{\itemsep}{0mm}
\bibitem[Bowman(1984)]{Bowman.1984}
Bowman, A. (1984). An alternative method of cross-validation for the smoothing of density estimates. \textit{Biometrika}, {\bf{71}}(2), pp.353-360.
\bibitem[Chen(2016)]{Chen.2016}
Chen, Y. and Samworth, R. J. (2016). Generalised additive and index models with shape constraints. \textit{Journal of the Royal Statistical Society: Series B}, {\bf{78}}(4), pp.729-754.
\bibitem[Forrest(1993)]{Forrest.1993}
Forrest, S. (1993).
Genetic Algorithms: Principles of natural selection applied to computation, 
\textit{Science}, {\bf{261}}(5123), pp.872-878.
\bibitem[Friedman(1984)]{Friedman.1984}
Friedman, J.H., Stuetzle, W., and Schroeder, A. (1984).
Projection pursuit density estimation, 
\textit{Journal of the American Statistical Association}, {\bf{79}}(387), pp.599-608.
\bibitem[Girolami(2003)]{Girolami.2003}
Girolami, M., and He, C. (2003). Probability density estimation from optimally condensed data samples,
\textit{IEEE Transactions on Pattern Analysis and Machine Intelligence}, {\bf{25}}(10), pp.1253-1264.
\bibitem[Goldberg(1988)]{Goldberg.1988}
Goldberg, D.E., and Holland, J.H. (1988).
Genetic algorithms and machine learning,
\textit{Machine Learning}, {\bf{3}}, pp.95-99.
\bibitem[Goldberg(1989)]{Goldberg.1989}
Goldberg, D.E. (1989). Genetic algorithms in search, optimization, and machine learning.
\textit{Addison-Wesley}.
\bibitem[Hastie(1990)]{Hastie.1990}
Hastie, T., and Tibshirani, R. (1990). Generalized additive models. \textit{Chapman and Hall}.
\bibitem[Haupt(2004)]{Haupt.2004}
Haupt, R.L., and Haupt, S.E.(2004). Practical genetic algorithms, Second edition. \textit{Wiley}.
\bibitem[He(2004)]{He.2004}
He, C., and Girolami, M.(2004). Novelty detection employing an L2 optimal non-parametric density estimator, \textit{Pattern Recognition Letters}, {\bf{25}}(12), pp.1389-1397.
\bibitem[Holland(1992)]{Holland.1992}
Holland, J.H. (1992).
Genetic algorithms, \textit{Scientific American}, {\bf{267}}(1), pp.66-72.
\bibitem[Huber(1985)]{Huber.1985}
Huber, P. J. (1985). Projection pursuit. {\textit{The Annals of Statistics}}, {\bf{13}}(2), pp.435-475.
\bibitem[Kessy(2018)]{Kessy.2018}
Kessy, A., Lewin, A., and Strimmer, K. (2018). Optimal whitening and decorrelation. \textit{The American Statistician}, {\bf{72}}(4), pp.309-314.
\bibitem[Liu(2011)]{Liu.2011}
Liu, H., Xu, M., Gu, H., Gupta, A., Lafferty, J., and Wasserman, L. (2011), Forest density estimation.
\textit{Journal of Machine Learning Research}, {\bf{12}}, pp.907-951.
\bibitem[Nishida(2023)]{Nishida.2023}
Nishida, K. (2023). Kernel density estimation by genetic algorithm,
\textit{Journal of Statistical Computation and Simulation}, {\bf{93}}(8), pp.1263-1281.
\bibitem[Nishida(2026)]{Nishida.2026}
Nishida, K. (2026). Sparse approximation of kernel density estimators via a genetic algorithm: data-size compression,
\textit{Kokyuroku}, {\bf{2336}}, pp.168-178.
%
\bibitem[Pagan(1999)]{Pagan.1999}
Pagan, A., and Ullah, A. (1999). Nonparametric econometrics. \textit{Cambridge press}.
\bibitem[Rudemo(1982)]{Rudemo.1982}
Rudemo, M. (1982) Empirical choice of histograms and kernel density estimators. \textit{Scandinavian Journal of Statistics}, {\bf{9}}, pp.65-78.
\bibitem[Schwaighofer(2007)]{Schwaighofer.2007}
Schwaighofer, A., Dejori, M., Tresp, M., and Stetter, M. (2007).
Structure learning with nonparametric decomposable models,
In J.M. de S\'a et al., \textit{Artificial Neural Networks-ICANN 2007, Lecture Notes in Computer Science}, {\bf{4668}}, pp.119-128.
\bibitem[Wand(1993)]{Wand.1993}
Wand, M.P., and Jones, M.C. (1993).
Comparison of smoothing parametrizations in bivariate kernel density estimation,
\textit{Journal of the American Statistical Association}, {\bf{88}}(422), pp.520-528.
\bibitem[Wand(1995)]{Wand.1995}
Wand, M.P., and Jones, M.C. (1995).
Kernel smoothing, \textit{Chapman and Hall}.
\bibitem[Zhang(2023)]{Zhang.2023}
Zhang, J., Meng, C., Yu, J., Zhang, M., Zhong, W, and Ma, P. (2023).
An optimal transport approach for selecting a representative subsample with application in efficient kernel density estimation, \textit{Journal of Computational and Graphical Statistics}, {\bf{32}}(1), pp.329-339.
\end{thebibliography}
\end{document}